\documentclass[twocolumn,floatfix,superscriptaddress,preprintnumbers,nofootinbib,10pt]{revtex4}
\usepackage{graphicx}
\usepackage{color}

\usepackage{amssymb,amsmath}

\newcommand{\bea}{\begin{eqnarray}}
\newcommand{\beq}{\begin{equation}}
\newcommand{\eea}{\end{eqnarray}}
\newcommand{\eeq}{\end{equation}}

\newcommand{\lsim}{\raise0.3ex\hbox{$\;<$\kern-0.75em\raise-1.1ex\hbox{$\sim\;$}}}
\newcommand{\gsim}{\raise0.3ex\hbox{$\;>$\kern-0.75em\raise-1.1ex\hbox{$\sim\;$}}}

\newcommand{\unity}{{\hbox{1\kern-.8mm l}}}

\newcommand{\lsp}{ \left ( }
\newcommand{\rsp}{ \right ) }

\newcommand{\mgut}{ M_{\rm G} }
\newcommand{\mpl}{ M_{\rm P} }

\newcommand{\cH}{{\cal H}}

\newcommand{\gev}{{\rm GeV}}
\newcommand{\epsK}{\varepsilon_K}

\begin{document}
\preprint{TUM-HEP-779/10}

\title{Footprints of SUSY GUTs in Flavour Physics}
\author{Andrzej~J.~Buras}
\affiliation{Physik-Department, Technische Universit\"at M\"unchen,
James-Franck-Stra\ss e, D-85748 Garching, Germany}
\affiliation{TUM Institute for Advanced Study, Technische Universit\"at M\"unchen,
\\Lichtenbergstr.~2a, D-85748 Garching, Germany}
\author{Minoru~Nagai}
\affiliation{Excellence Cluster Universe, Technische Universit\"at M\"unchen,
D-85748 Garching, Germany}
\author{Paride~Paradisi}
\affiliation{Physik-Department, Technische Universit\"at M\"unchen,
James-Franck-Stra\ss e, D-85748 Garching, Germany}

\begin{abstract}
Supersymmetric (SUSY) Grand Unified theories (GUTs) generally predict FCNC and CP violating 
processes to occur both in the leptonic and hadronic sectors. Assuming an underlying $SU(5)$
group plus right-handed neutrinos (RN), we perform an extensive study of FCNC and CP violation,
analyzing the correlations between leptonic and hadronic processes like $\mu\to e\,\gamma$ and $K^0-\bar{K}^0$ mixing, $\tau\to\mu\,\gamma$ and $b \to s$ transitions such as $B_d\to\phi K_s$
and $B^{0}_s-\bar{B}^{0}_s$ mixing.
Moreover, we examine the impact of the considered scenario on the UT analyses, monitoring the
low energy consequences implied by possible solutions to the various tensions in the present
UT analyses. We compare the phenomenological implications of this NP scenario with the ones
of supersymmetric flavour models finding a few striking differences that could allow to
distinguish these different NP models.

\end{abstract}

\maketitle

\section{Introduction}

Grand unified theories (GUTs)~\cite{GUT} predict a very attractive unification of the
strong and electroweak interactions and, when embedded in a supersymmetric framework,
they also lead to a very successful gauge coupling unification~\cite{gauge_coupl_unif}.

Since a direct access to the GUT scale ($M_G$) is not possible, experimental signals
for GUTs may be looked for either by confronting their predictions for quantities that
are not predicted in the SM, such as the weak mixing angle and the $m_b/m_\tau$ ratio
or by the observation of processes which are forbidden or highly suppressed in the
SM, such as the proton decay.
In particular, the baryon number $B$, the family lepton numbers $L_e, L_\mu, L_\tau$ and
the total lepton number $L = L_e+L_{\mu}+L_{\tau}$, which are accidental symmetries of 
the SM, are broken in a GUT context.
As a result, processes violating $B$, $L$ or $L_i$ are generated and they turn out to be
suppressed by powers of $1/M_{G}$. Still, the experimental sensitivity to proton
decay and to neutrino masses (assuming an underlying see-saw mechanism~\cite{seesaw})
allow to probe $M_{G}$. Indeed, the observed neutrino masses and mixings give solid
indications for the existence of some New Physics (NP) at $M_{G}$. In contrast,
$L_i$ violating processes, such as $\mu \to e\gamma$, are predicted to be far below
any realistic experimental sensitivity.

If the theory above $M_{G}$ is supersymmetric, the situation drastically changes.
The dynamics at the scale $M_{G}$ leave indelible traces on the soft terms of the
light sparticles by means of interactions not suppressed by inverse powers of
$M_{G}$~\cite{hkr,borzumatimasiero}.
Processes which violate $L_i$ are now suppressed only by powers of $1/{\tilde m}$, where
${\tilde m}$ is the scale of supersymmetry breaking, and therefore they might be measurable.

Moreover, since within GUTs quarks and leptons sit in the same multiplets, the quark-lepton
unification feeds into the SUSY breaking soft sector~\cite{hkr,barbieri,correlations}.

This implies relations between lepton and quark flavor changing transitions at the
weak scale. For example, one can expect correlations between $\mu\to e\,\gamma$ and $K^0-\bar{K}^0$ mixing, $\tau\to\mu\,\gamma$ and $b \to s$ transitions such as
$B_d\to\phi K_s$ and $B^{0}_s-\bar{B}^{0}_s$ mixing and so on. Therefore, hadronic
and leptonic FCNC processes provide a splendid opportunity to link the weak scale
to the GUT scale, where the fundamental SUSY Lagrangian is defined.

Despite of the remarkable agreement of flavour data with the SM predictions in the $K$
and $B_d$ systems, a closer look at the data might indicate some tensions especially in
CP violating observables. In particular:

i) the most recent UT analyses~\cite{SL,Buras:2008nn,Lenz:2010gu,Bevan:2010gi} show that
the size of CP violation determined through the $B^{0}_d-\overline{B}^{0}_d$ system,
appears insufficient to describe the experimental value of $\epsilon_K$ within the SM
if the $\Delta M_d/\Delta M_s$ constraint is taken into account~\cite{Buras:2008nn}.
Vice versa, the simultaneous SM description of $\epsilon_K$ and $\Delta M_d/\Delta M_s$
requires a $\sin 2\beta$ significantly larger than the measured value of
$S_{\psi K_S}$~\cite{SL}.

ii) The recent messages from the Tevatron seem to hint the presence of new sources of
CPV entering the $B^{0}_s$ system~\cite{Aaltonen:2007he,Abazov:2010hv,Abazov:2008fj}.

iii) The value of $\sin 2\beta$ extracted from some penguin dominated modes, such as
$B_d\to\phi K_S$, is significantly lower than the value from $B_d\to\psi K_S$.

iv) Last but not least, we remind the $(g-2)_{\mu}$ anomaly~\cite{g_2_th}. Interestingly,
the possibility that the present $3\sigma$ discrepancy may arise from errors in the
determination of the hadronic leading-order contribution to $\Delta a_{\mu}$ seems
to be unlikely~\cite{passera_mh}.

In the light of the above considerations, in this work we focus on the SUSY $SU(5)$
GUT model plus right-handed neutrinos~\cite{Hisano:1997tc} ($SSU(5)_{RN}$),
accounting for the neutrino masses and mixings via a type-I sees-saw model~\cite{seesaw},
with the main goals:
\begin{itemize}
\item[\bf i)] to analyze how the $SSU(5)_{RN}$ model faces the above tensions,
monitoring the low energy consequences implied by their possible solutions;
\item[\bf ii)] to quantify the NP room left for $b\to s$ transitions compatible
with all the available experimental data on $\Delta F=2$ and $\Delta F=1$ processes,
\item[\bf iii)] to outline strategies aimed to probe or to falsify the $SSU(5)_{RN}$
model by means of a correlated analysis of low energy observables, including also
$K$ and $D$ systems as well as lepton flavour violation, electric dipole moments
(EDMs) and the $(g-2)_{\mu}$.
\end{itemize}

Since many analyses along this subject appeared in the literature~\cite{Parry:2007fe},
we want to emphasize here that the current study goes well beyond previous works as for
i) the inclusion of all relevant SUSY contributions, ii) the number of processes considered,
and iii) the special attention given to the UT analyses. Moreover, we point out many new
correlations among observables that should enable us to probe or falsify the scenario in
question once improved data will be available.

In Section 2 we update our UT analysis of~\cite{Altmannshofer:2009ne} using the $(R_b,\gamma)$
plane and stressing that a large value of $\gamma$ in the UT would provide a natural solution
to the observed tensions within the $SSU(5)_{RN}$ model. In Section 3 we summarize the flavour
structure of the SUSY GUT considered in this paper. In Sections 4 and 5 the relevant formulae
for the hadronic and the leptonic sectors are given, respectively.
Section 6 is devoted to approximate analytical expressions for various correlations between
hadronic and leptonic observables that are then analyzed numerically in Section 7.
In Section 8 we present a DNA-table for this GUT scenario and compare it with the ones of supersymmetric flavour models analyzed by us in~\cite{Altmannshofer:2009ne}. Finally we end
our paper with a list of the most important findings.
\begin{table*}
\begin{center}
\begin{tabular}{|l|l||l|l|}
\hline
parameter & value & parameter & value \\
\hline\hline
$F_K$ & $(155.8\pm 1.7) \text{MeV}$	\cite{Laiho:2009eu,Antonio:2007pb}  & $m_s(2\,\text{GeV})$&$0.105\,\text{GeV}$~\cite{Amsler:2008zzb}\\
$F_{B_d}$ & $(192.8 \pm 9.9) \text{MeV}$ \cite{Laiho:2009eu,Antonio:2007pb} 	  & $m_d(2\,\text{GeV})$&$0.006\,\text{GeV}$~\cite{Amsler:2008zzb}\\
$F_{B_s}$ & $(238.8 \pm 9.5) \text{MeV}$\cite{Laiho:2009eu,Antonio:2007pb}	  & $|V_{ts}|$&$0.040\pm 0.003$~\cite{Bona:2007vi}\\
$\hat B_K$ & $0.725 \pm 0.026$ 	\cite{Laiho:2009eu,Antonio:2007pb}		  & $|V_{tb}|$&$ 1\pm 0.06$~\cite{Bona:2007vi}\\
$\hat B_{B_d}$ & $1.26\pm 0.11$ \cite{Laiho:2009eu,Antonio:2007pb}		  & $|V_{td}|_{\rm tree}$&$(8.3\pm 0.5)\cdot 10^{-3}$~\cite{Bona:2007vi}\\
$\hat B_{B_s}$ & $1.33\pm 0.06$	\cite{Laiho:2009eu,Antonio:2007pb}	          & $|V_{us}|$&$ 0.2255 \pm 0.0019 $\cite{Amsler:2008zzb}\\
$M_{B_s}$&$5.3664$ GeV~\cite{Amsler:2008zzb}                       & $|V_{cb}|$&$ (40.6 \pm 1.1)\times 10^{-3}$\cite{Nakamura:2010zzi}\\
$M_{B_d}$&$5.2795$ GeV~\cite{Amsler:2008zzb}                       & $\sin(2\beta)_{\rm tree}$ & $0.734\pm 0.038$~\cite{Bona:2007vi}\\
$M_K$&$0.497614\,\text{GeV}$~\cite{Amsler:2008zzb}		  & $\sin(2\beta_s)$&$0.038\pm 0.003$~\cite{Bona:2007vi}\\ 
$\eta_{cc}$ & $1.43\pm 0.23$\cite{Herrlich:1996vf}		  & $\alpha_s(m_Z)$&$ 0.1184$~\cite{Bethke:2009jm}\\
$\eta_{tt}$ & $0.5765\pm 0.0065$\cite{Buras:1990fn}		  & $\Delta M_s$ & $(17.77\pm 0.12)~{\rm ps}^{-1}$ ~\cite{Barberio:2008fa}\\
$\eta_{ct}$ & $0.496\pm 0.047$\cite{Brod:2010mj}		  & $\Delta M_d$ & $(0.507\pm 0.005)~ {\rm ps}^{-1}$ ~\cite{Barberio:2008fa}\\
$\eta_{B}$  & $0.551\pm 0.007$\cite{Buras:1990fn}  & $\Delta M_K$&$(5.292\pm 0.009)\cdot 10^{-3} ps^{-1}$~\cite{Amsler:2008zzb}\\
$\xi$ & $1.243 \pm 0.028$\cite{Laiho:2009eu,Antonio:2007pb}	                  & $\kappa_\varepsilon$& $0.94\pm 0.02$~\cite{Buras:2010pz} \\
$m_c(m_c)$ & $(1.268\pm 0.009) \gev$\cite{Allison:2008xk}	  & $\epsK^{\text{exp}}$&$(2.229\pm 0.01)\cdot 10^{-3}$~\cite{Amsler:2008zzb}\\
$m_t(m_t)$ & $(163.7\pm 1.1) \gev$\cite{CDF_D0_mt}		  & $S_{\psi K_S}^{\text{exp}}$& $0.672\pm0.023$~\cite{Barberio:2008fa}\\
$m_b(m_b)$& $ (4.2+0.17-0.07) \gev$~\cite{Amsler:2008zzb}          & & \\
\hline
\end{tabular}
\caption{Values of the input parameters used in our analysis. The subscript ``tree''
in $|V_{td}|$ and $\sin(2\beta)$ stands for the inputs extracted from data using only
tree-level observables~\cite{Bona:2007vi}.}
\label{tab:eps}
\end{center}
\end{table*}
\begin{figure*}[th]
\includegraphics[width=0.4\textwidth]{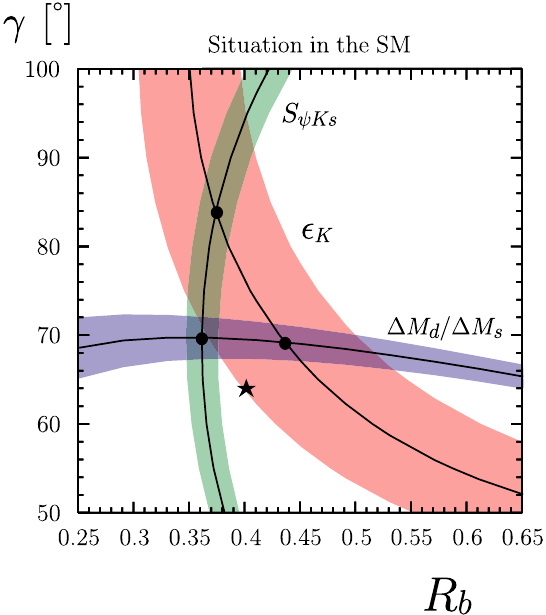}~~
\includegraphics[width=0.4\textwidth]{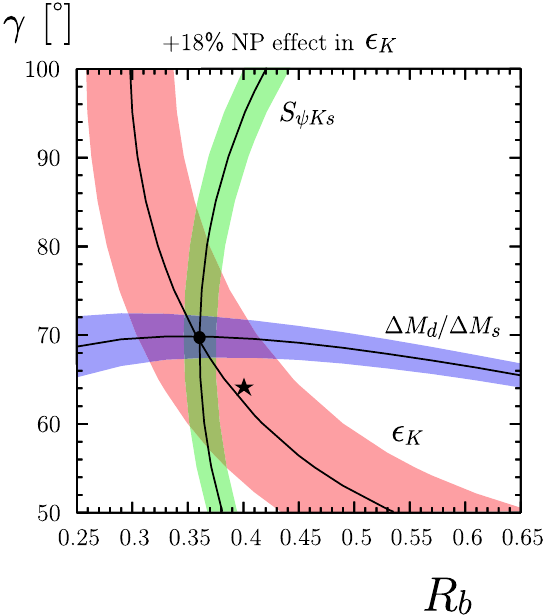}\\
\includegraphics[width=0.4\textwidth]{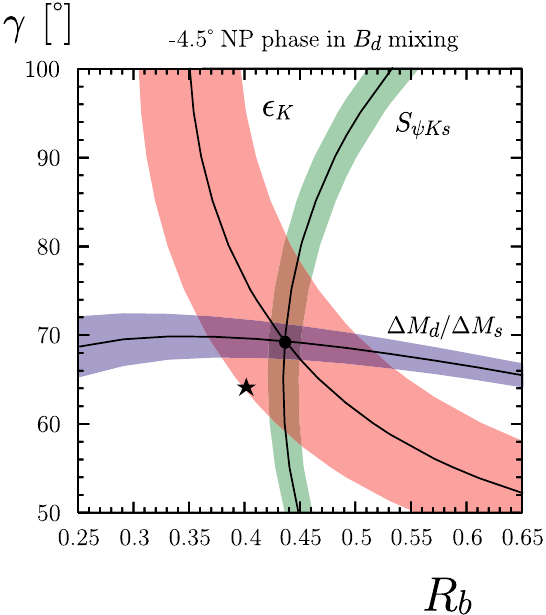}~~
\includegraphics[width=0.4\textwidth]{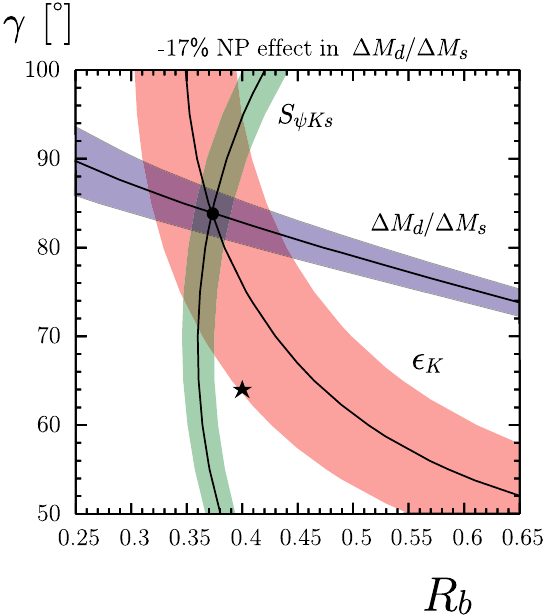}
\caption{
The $R_b-\gamma$ plane assuming: {\bf i)} the SM situation with the input parameters of
Table~\ref{tab:eps} (upper left), {\bf ii)} $\sin 2\beta$ and $R_t$ NP free while $\epsilon_K$
affected by a $+18\%$ NP effect compared to the SM contribution (upper right), {\bf iv)}
$\epsilon_K$ and $R_t$ NP free while $\sin 2\beta$ affected by a NP phase in $B_d$ mixing
with $\varphi_{B_d}\approx -4.5^\circ$ (lower left), {\bf iii)} $\epsilon_K$ and $\sin 2\beta$
NP free while $\Delta M_d/\Delta M_s$ affected by a $-20\%$ NP contribution compared to the 
SM contribution (lower right). The black star stands for the values obtained from the NP UT fit~\cite{Bona:2005eu}.}
\label{fig1}
\end{figure*}

%%%%%%%%%%%%%%%%%%%%%%%%%%%%%%%%%%%%%%%%%%%%%%%%%%%%%%%%%%%%%%%%%%%%%%%%%%%%%%%%%%%%%%%%%%%%%%
\section{UT analysis} \label{sec:UT_analysis}
%%%%%%%%%%%%%%%%%%%%%%%%%%%%%%%%%%%%%%%%%%%%%%%%%%%%%%%%%%%%%%%%%%%%%%%%%%%%%%%%%%%%%%%%%%%%%%
%
In this section, we perform a unitarity triangle (UT) analysis in the framework of the
$SSU(5)_{RN}$ model. We remind that there exist two different UTs: 1) the so-called 
reference unitarity triangle (RUT)~\cite{Goto:1995hj}, determined entirely from tree level
decays hence, likely unaffected by any significant NP pollution, and 2) the universal 
unitarity triangle (UUT)~\cite{Buras:2000dm} of models with constrained MFV, determined
by means of loop-induced FCNC processes and hence potentially sensitive to NP effects.
Therefore, a comparative UT analysis performed by means of the RUT and UUT may unveil 
NP effects.
In particular, the above UTs are characterized by the following parameters
\bea
\label{eq:UT_parameter_1}
V_{us} &\equiv&
\lambda ~,~ V_{cb} ~,~ R_b ~,~ \gamma \qquad {\rm RUT}\,,
\nonumber\\
\label{eq:UT_parameter_2}
V_{us} &\equiv&
\lambda ~,~ V_{cb} ~,~ R_t ~,~ \beta  \qquad {\rm UUT}\,,
\eea
where $R_b\equiv|V_{ud}V^*_{ub}|/|V_{cd}V^*_{cb}|$, $R_t\equiv|V_{td}V^*_{tb}|/|V_{cd}V^*_{cb}|$
and the angles $\beta$ and $\gamma$ are such that $V_{td}=|V_{td}|e^{-i\beta}$ and $V_{ub}=|V_{ub}|e^{-i\gamma}$. Moreover, the dictionary between $(R_t,\beta)$ and
$(R_b,\gamma)$ reads
\begin{equation} \label{eq:Rt_beta}
R_t\!=\!\sqrt{1+R_b^2-2 R_b\cos\gamma} ~,~~
\cot\beta\!=\!\frac{1-R_b\cos\gamma}{R_b\sin\gamma}~,
\end{equation}
\begin{equation} \label{eq:Rb_gamma}
R_b\!=\!\sqrt{1+R_t^2-2 R_t\cos\beta} ~,~~
\cot\gamma\!=\!\frac{1-R_t\cos\beta}{R_t\sin\beta}~.
\end{equation}
In terms of physical observables we can write
\begin{equation}
\label{eq:Rt_sin2beta_NP}
R_t = \frac{\xi}{\lambda} \sqrt{\frac{m_{B_s}}{m_{B_d}}} \sqrt{\frac{\Delta M_d}{\Delta M_s}} \sqrt{\frac{C_{B_s}}{C_{B_d}}} ~,~~~  \sin(2 \beta + 2 \varphi_{B_d}) = S_{\psi K_S} ~,
\end{equation}
with the SM limit recovered for $C_{B_q}=1$ and $\varphi_{B_d}=0$.

The last observable that is relevant for our UT analysis is $\epsilon_K$.
In the SM, $\epsilon_K$ can be written as~\cite{Buras:2008nn}
\bea
\label{eq:epsK_SM}
|\epsilon_K|\!&=&\!
\kappa_\epsilon C_\epsilon\hat B_K |V_{cb}|^2 |V_{us}|^2
\bigg(\frac{|V_{cb}|^2}{2} R_t^2 \sin 2\beta \eta_{tt}S_0(x_t)+
\nonumber\\
&+&R_t\sin\beta(\eta_{ct}S_0(x_c,x_t)-\eta_{cc}x_c)\bigg),
\eea
where $C_\epsilon \simeq3.658\times 10^4$ and all the parameters entering the above
expression are reported in Table~\ref{tab:eps}.
As stressed in~\cite{Buras:2008nn}, the SM prediction for $\epsilon_K$ implied by the
measured value of $S_{\psi K_S}=\sin2\beta$ may be too small to agree with experiment.
The main reasons are the decreased value of $\hat B_K$ and the decreased value of
$\epsilon_K$ in the SM arising from a multiplicative factor, estimated as
$\kappa_\epsilon=0.94\pm 0.02$~\cite{Buras:2010pz}.

Taking into account also the recent calculation of the QCD factor $\eta_{ct}$ at the 
NNLO~\cite{Brod:2010mj}, that enhances the value of $\varepsilon_K$ by $3\%$~\cite{Brod:2010mj},
the total suppression of $\epsilon_K \propto\hat B_K \kappa_\epsilon$ compared to the 
commonly used formulae is typically of order 15\%. Using the inputs of Table~\ref{tab:eps}, eq.~(\ref{eq:Rt_sin2beta_NP}) for the SM case (where $C_{B_s}=C_{B_d}=1$), and
eq.~(\ref{eq:epsK_SM}), one finds~\cite{Brod:2010mj}
\begin{equation} \label{eq:epsK_SMnumber}
|\epsilon_K|^{\rm SM} = (1.90 \pm 0.26) \times 10^{-3}~,
\end{equation}
to be compared with the experimental measurement \cite{Amsler:2008zzb}
\begin{equation} \label{eq:epsK_exp}
|\epsilon_K|^{\rm exp} = (2.229 \pm 0.010) \times 10^{-3}~.
\end{equation}
In fig.~\ref{fig1}, we show the above tensions in the $R_b-\gamma$ plane updating
the analysis of~\cite{Altmannshofer:2009ne} by the inclusion of the new values for $\kappa_\epsilon$~\cite{Buras:2010pz} and $\eta_{ct}$~\cite{Brod:2010mj}.

In the upper left plot of fig.~\ref{fig1}, we show the regions corresponding to the 1$\sigma$
allowed ranges for $\sin2\beta$, $R_t$ and $|\epsilon_K|^{\rm SM}$ as calculated by means of~(\ref{eq:Rt_sin2beta_NP}) and (\ref{eq:epsK_SM}), respectively, using the numerical
input parameters of tab.~\ref{tab:eps}. As shown, there are three different values of
$(R_b,\gamma)$, dependently which two constraints are simultaneously applied.

Possible solutions to this tension can be obtained assuming:
\begin{itemize}
\item[\bf 1)] a positive NP contribution to $\epsilon_K$, at the level of $\approx +20\%$,
leaving $\sin 2\beta$ and $\Delta M_d/\Delta M_s$ SM-like~\cite{Buras:2008nn}.
\item[\bf 2)] $\epsilon_K$ and $\Delta M_d/\Delta M_s$ NP free while $S_{\psi K_S}$
affected by a NP phase in $B_d$ mixing with $\varphi_{B_d}\approx -5^\circ$~\cite{SL}.
\item[\bf 3)] $\epsilon_K$ and $S_{\psi K_S}$ NP free while $\Delta M_d/\Delta M_s$
affected by NP at the level of $\approx -20\%$~\cite{Altmannshofer:2009ne},
requiring in turn an increased value of $R_t$ to fit the data.
\end{itemize}
Of course all these effects could be simultaneously at work.

As stressed in~\cite{Altmannshofer:2009ne}, the possibility 3) implies a large value of
$\gamma$, as shown in fig.~\ref{fig1} (see the lower plot on the right), and $\alpha$
significantly below $90^\circ$. Therefore, this scenario can be easily probed or falsified
by means of improved determinations of $\gamma$, as expected at the LHCb, and $\alpha$.
Moreover, the possibility 3) is particularly relevant within the $SSU(5)_{RN}$ model,
as we will discuss in detail later. In such a case, if the mixing angle regulating the
$b\to s$ transition contains a natural $\mathcal{O}(1)$ CPV phase, then solution 3)
also implies a non-standard value for $S_{\psi\phi}$ in the $B_s^0$ system.

\section{SUSY GUTs and flavour phenomenology}

The quark-lepton unification predicted by GUTs implies, in a SUSY framework,
a unification of the squark and slepton mass matrices at the GUT scale.

For instance, within SU(5), the multiplet ${\bar {\bf 5}}$ contains the right-handed
down-type quarks ($D^c$) and the left-handed lepton doublets ($L$) while the $\bf 10$
multiplet contains the left-handed quark doublets ($Q$), the right-handed up-type 
quarks ($U^c$), and the right-handed charged-leptons ($E^c$).

Even if we assume a complete flavour blindness for the sfermion masses at the high
scale (either $M_P$ or $M_G$), we can still expect sizable sources of FCNC if the
theory contains neutrino Yukawa interactions accounting for the neutrino masses and
mixings by means of a see-saw mechanism.

Assuming a type-I see-saw with three heavy right-handed neutrinos~\cite{seesaw}, the
effective light-neutrino mass matrix resulting after integrating out the heavy fields
is $m_\nu \!=\! y_\nu^T\hat{M}_\nu^{-1} y_\nu \langle H_u \rangle^2$ where $\hat{M}_\nu$
is the right-handed neutrino mass matrix, $y_{\nu}$ is the unknown neutrino Yukawa
matrix, and $\langle H_u \rangle$ is the up-type Higgs VEV. Hereafter, we take a basis
where $\hat{M}_\nu$ is diagonal and symbols with hat mean they are diagonal matrices.

In the mSUGRA scenario and specializing to the case of $SSU(5)_{RN}$ model, low-energy flavor-violating SUSY-breaking terms are radiatively induced, and they are qualitatively
given as
\bea
 (m_{{\tilde d}_R}^2)_{ij}\!\!&=&\!\!
- \frac{(3m_0^2\!+\!A_0^2)}{8\pi^2}
    (e^{i\hat{\phi}_{d}} y^T_\nu y^*_\nu
    e^{-i\hat{\phi}_{d}})_{ij}\ln \frac{\mpl}{\mgut},
\nonumber\\
 (m_{{\tilde d}_L}^2)_{ij}\!\!&=&\!\!
- \frac{(3m_0^2\!+\!A_0^2)}{8\pi^2} (V^\dagger \hat{y}_u^2 V)_{ij}
    \lsp 3\ln \frac{\mpl}{\mgut} \!+\! \ln \frac{\mgut}{\tilde m} \rsp,
\nonumber\\
(m_{{\tilde e}_R}^2)_{ij}\!\!&=&\!\!
-3\frac{(3m_0^2\!+\!A_0^2)}{8\pi^2}
(e^{i\hat{\phi}_d} V^T \hat{y}_u^2 V^* e^{-i\hat{\phi}_d})_{ij}
\ln\frac{M_{\rm P}}{M_{\rm G}}\,,
\nonumber\\
 (m_{{\tilde l}_L})_{ij}\!\!&=&\!\!
 - \frac{(3m_0^2\!+\!A_0^2)}{8\pi^2}
 (y^{\dagger}_{\nu})_{ik} (y_{\nu})_{kj}\ln\frac{\mpl}{M_{\nu_k}},
\label{Eq:SU5RN_FV}
\eea
where $V$ is the CKM matrix, $\hat{\phi}_d$ is a GUT phase and $m_0$ ($A_0$) is the
universal scalar mass (trilinear coupling).

Within $SU(5)$, as both $Q$ and $E^c$ are hosted in the ${\bf 10}$ representation, the CKM
matrix mixing of the left-handed quarks will give rise to off-diagonal entries in the running
of the right-handed slepton soft masses $(m_{{\tilde e}_R}^2)_{ij}$ due to the interaction
of the colored Higgs~\cite{hkr,barbieri}. Vice versa, $Y_{\nu}$ enters the mass matrices
of both right-handed down squark and left-handed sleptons, as $D^c$ and $L$ lie in the
${\bar {\bf 5}}$ multiplet of $SU(5)$.

We remind that, within a type-I see-saw scenario, $y_\nu$ can be written in the general form~\cite{Casas:2001sr} $y_\nu=\sqrt{\hat{M}_\nu}R\sqrt{\hat{m}_\nu}U^{\dagger}/\langle H_u\rangle$
where $R$ is an arbitrary complex orthogonal matrix while $U$ is the PMNS matrix.
The determination of $(m_{{\tilde{l}_L}}^2)_{i\neq j}$ would require a complete knowledge
of the neutrino Yukawa matrix $y_\nu$, which is not possible using only low-energy
observables from the neutrino sector.
In particular, the ratio $(m_{{\tilde{l}_L}}^2)_{12}/(m_{{\tilde{l}_L}}^2)_{23}$ is highly
model dependent while $(m_{{\tilde{l}_L}}^2)_{12}\simeq(m_{{\tilde{l}_L}}^2)_{13}$ to a good approximation~\cite{Hisano:2009ae}. The situation is completely different in a
type-II see-saw scenario where the mixing angles of $(m_{{\tilde{l}_L}}^2)_{i\neq j}$ are
entirely governed by the PMNS matrix and low energy LFV processes relative to different
family transitions are strikingly correlated~\cite{Rossi:2002zb}.

Furthermore, we remind that a successful Yukawa coupling unification requires either the
introduction of some GUT breaking effects or some new non-renormalizable contributions.
In both cases, we are forced to introduce a relative rotation matrix $V^{(ql)}$ between
the quark and leptonic fields (defined in the super-CKM basis) such that the quark-lepton
correlations might be modified~\cite{Borzumati:2009hu}. However, since the matrix $V^{(ql)}$
is unknown, in the following, we assume the case where the naive quark-lepton correlations
are still valid.

%%%%%%%%%%%%%%%%%%%%%%%%%%%%%%%%%%%%%%%%%%%%%%%%%%%%%%%%%%%%%%%%%%%%%%%%%%%%%%%%%%%%%%%%%
\section{Hadronic sector}
%%%%%%%%%%%%%%%%%%%%%%%%%%%%%%%%%%%%%%%%%%%%%%%%%%%%%%%%%%%%%%%%%%%%%%%%%%%%%%%%%%%%%%%%%

In what follows, we discuss the relevant observables for our study, including
$\Delta F =2,1,0$ transitions.

{\bf 1.} The complete set of operators for $\Delta B=2$ transitions is~\cite{Buras:2001ra}
\bea
Q_1^{VLL}&=&(\bar b_L \gamma_\mu q_L)(\bar b_L \gamma^\mu q_L)\,, \nonumber\\
Q_1^{SLL}&=&(\bar b_R q_L)(\bar b_R q_L)\,, \nonumber\\
Q_2^{SLL}&=&(\bar b_R \sigma_{\mu\nu} q_L)(\bar b_R \sigma^{\mu\nu} q_L)\,, \nonumber\\
Q_1^{LR}&=&(\bar b_L \gamma_\mu q_L)(\bar b_R \gamma^\mu q_R)\,, \nonumber\\
Q_2^{LR}&=&(\bar b_R q_L)(\bar b_L q_R)\,,
\label{eq:operatorsDF2}
\eea
where $q=s,d$, $\sigma_{\mu\nu}=\frac{1}{2}\left[\gamma_\mu,\gamma_\nu\right]$.
In Eq.~(\ref{eq:operatorsDF2}) we did not show the operators $Q_1^{\rm VRR}$ and
$Q_{1,2}^{SRR}$ that are obtainable from $Q_1^{\rm VLL}$ and $Q_{1,2}^{SLL}$,
respectively, with the exchange $q_L\to q_R$.

The low energy effective Hamiltonian reads~\cite{Buras:2001ra}
\beq
\cH_{\rm eff}=\sum_{i,a}C_i^a(\mu_B,B)Q_i^a~,
\eeq
where the summation is performed over contributing operators. The off-diagonal
element in the $B^{0}_q$-meson mixing is given by
\beq
\label{eq:M12q}
M_{12}^q = \frac{1}{3}M_{B_q} F_{B_q}^2\sum_{i,a} C_i^{a*}(\mu_H,B_q) P_i^a(B_q)\,,
\eeq
where $P_i^a(B^{0}_q)$ collect all RG effects from the high scale, where heavy degrees of freedom
are integrated out, down to the $B$ meson scale as well as hadronic matrix elements obtained
by lattice QCD techniques. Updating the results of Ref.~\cite{Buras:2001ra}, it turns out that
$P_2^{LR} (B^{0}_q)\approx 3.4$ and $P_1^{SLL} (B^{0}_q)\approx -1.4$ for $\mu_H=246$~GeV.

Introducing the notation
\begin{equation}
M_{12}^q=\left(M_{12}^q\right)_{\text{SM}} C_{B_q}e^{2 i\varphi_{B_q}}~,
\qquad (q=d,s)~,
\label{eq:M12}
\end{equation}
the $B^{0}_{s,d}$ mass differences and the CP asymmetries $S_{\psi K_S}$ and $S_{\psi\phi}$ are
\bea
\Delta M_q &=& 2\left|M_{12}^q\right| = (\Delta M_q)_{\text{SM}}C_{B_q}~,
\label{eq:Delta_Mq} \\
S_{\psi K_S} &=& \sin( 2\beta + 2\varphi_{B_d} )~,\\
S_{\psi\phi} &=& \sin( 2|\beta_s| - 2\varphi_{B_s} )~,
\label{eq:CPV_Bq}
\eea
where $\sin(2\beta)_{\rm tree}=0.734\pm 0.038$~\cite{Bona:2007vi} and
$\sin(2\beta_s)=0.038\pm 0.003$~\cite{Bona:2007vi}.

The $\Delta S=2$ operators are obtained from Eq.~(\ref{eq:operatorsDF2}) by means of $b\to s$
and $q=d$. The off-diagonal element in $K^0-\bar K^0$ mixing $M_{12}^K$ is then given by
\beq
 2 M_K M_{12}^K=\langle \bar K^0|\cH_{\rm eff}|K^0\rangle^*
\eeq
and the observables $\Delta M_K$ and $\epsilon_K$ can be evaluated through
\bea
\Delta M_K&=&2 \text{Re}\left( M_{12}^K\right)\,,\\
\epsilon_K&=&e^{i\varphi_\epsilon}\frac{\kappa_{\epsilon}}{\sqrt 2\, 
\Delta M_K}{\text{Im}}\left(M_{12}^K\right)\,,
\eea
where $\kappa_\varepsilon=0.94\pm0.02$~\cite{Buras:2008nn,Buras:2010pz} accounts
for $\varphi_\varepsilon=(43.51\pm0.05)^\circ\neq\pi/4$ and includes long distance
contributions.

%%%%%%%%%%%%%%%%%%%%%%%%%%%%%%%%%%%%%%%%%%%%%%%%%%%%%%%%%%%%%%%%%%%%%%%%%%%%%%%%%%%%%%%%%%%

Let us discuss now the semileptonic asymmetry in $B^{0}$ decays $A^{b}_{\text{SL}}$, described
by the quantity~\cite{Abazov:2010hv}
\begin{equation}
A^{b}_{\text{SL}} = (0.506 \pm 0.043) A^{d}_{\text{SL}} + (0.494 \pm 0.043) A^{s}_{\text{SL}}~,
\end{equation}
where $A^{d}_{\text{SL}}$ and $A^{s}_{\text{SL}}$ are the asymmetries in $B^0_d$ and $B^0_s$
decays, respectively. Within the SM, it is predicted that
$A^{d}_{\text{SL}}({\rm SM})=(-0.48 ^{+0.1}_{-0.12}) \times 10^{-3}$~\cite{Nierste},
$A^{s}_{\text{SL}}({\rm SM})=(2.1 \pm 0.6) \times 10^{-5}$~\cite{Nierste} and
therefore~\cite{Nierste}
\begin{equation}
A^{b}_{\text{SL}}({\rm SM}) = (-0.23^{+0.05}_{-0.06}) \times 10^{-3}~.
\label{aslbsm}
\end{equation}
This has to be compared with the recently measured value by the D0 collaboration~\cite{Abazov:2010hv}
\beq
A^{b}_{\text{SL}}({\rm D0}) = (-9.57 \pm 2.51 \pm 1.46)\times 10^{-3}~,
\label{ASL_exp}
\eeq
that differs by 3.2 standard deviations from the SM prediction, providing the first
evidence for anomalous CP-violation in the mixing of neutral $B^{0}$ mesons.

In the presence of NP, the asymmetries $A^{d}_{\text{SL}}$ and $A^{s}_{\text{SL}}$
can be evaluated by means of the following expression~\cite{Nierste}
\begin{equation} 
\label{eq:A_SL_corr}
A^{q}_{\text{SL}} =
{\rm Im}\left(\!\frac{\Gamma^{q}_{12}}{M^{q}_{12}}\!\right)^{\rm \!SM}
\!\frac{\cos 2\varphi_{B_q}}{C_{B_q}}-
{\rm Re}\left(\!\frac{\Gamma^{q}_{12}}{M^{q}_{12}}\!\right)^{\rm \!SM}
\!\frac{\sin 2\varphi_{B_q}}{C_{B_q}}~,
\end{equation}
where the updated values for ${\rm Re}(\Gamma^{q}_{12}/M^{q}_{12})^{\rm SM}$ and
${\rm Im}(\Gamma^{q}_{12}/M^{q}_{12})^{\rm SM}$ can be found in~\cite{Nierste}.
Moreover, we recall that $A^{s}_{\text{SL}}$, in the presence of NP and neglecting
$\beta_s$, is correlated model-independently with $S_{\psi\phi}$ as
$A^{s}_{\text{SL}}\simeq S_{\psi\phi}\!\times\!({\rm R}^{s}_{12}/C_{B_s})$~\cite{Ligeti:2006pm}
(for an alternative model-independent formula, see~\cite{Grossman:2009mn}).

%%%%%%%%%%%%%%%%%%%%%%%%%%%%%%%%%%%%%%%%%%%%%%%%%%%%%%%%%%%%%%%%%%%%%%%%%%%%%%%%%%%%%%%%%%%

On general grounds, we observe that the $A^{d}_{\text{SL}}$ contribution to
$A^{b}_{\text{SL}}$ is constrained model-independently by the limited NP room
left to $S_{\psi K_S}$. Furthermore, since in the $SSU(5)_{RN}$ model
$(m_{\tilde{l}_L}^{2})_{12}\simeq (m_{\tilde{l}_L}^{2})_{13}$, implying that $(m_{\tilde{d}_R}^{2})_{12}\simeq (m_{\tilde{d}_R}^{2})_{13}$, it turns out
that $S_{\psi K_S}$ is SM-like to a very good extent after imposing the experimental
bound on $\epsilon_K$. As a result we conclude that, within the $SSU(5)_{RN}$
model, the NP contributions to $A^{d}_{\text{SL}}$ are completely negligible.
By contrast, the NP contributions to $A^{s}_{\text{SL}}$ might be large since in
this case the limits from $S_{\psi K_S}$ and $\epsilon_K$ do not generally apply.

Therefore, within the $SSU(5)_{RN}$ model, it turns out that
\begin{equation}
\label{eq:Ab_SL_Y}
A^{b}_{\text{SL}} \approx 0.5~A^{s}_{\text{SL}}
\approx - 10^{-3}~\frac{S_{\psi\phi}}{C_{B_s}}\,,
\end{equation}
where $S_{\psi\phi}\lesssim 1$ and therefore $A^{b}_{\text{SL}}\gtrsim -10^{-3}$.

%%%%%%%%%%%%%%%%%%%%%%%%%%%%%%%%%%%%%%%%%%%%%%%%%%%%%%%%%%%%%%%%%%%%%%%%%%%%%%%%%%%%%%%%%%%%%%%

We briefly recall now the leading SUSY contributions to the $K^0$, $B^{0}_{d}$
and $B^{0}_{s}$ mixing amplitudes coming from gluino/squark boxes and, in the
large $\tan\beta$ regime, from double Higgs penguin contributions (we refer to ref.~\cite{Altmannshofer:2009ne} and therein references for the full expressions).
The dominant gluino/squark boxes read
\begin{eqnarray}
\label{eq:MIA2_C4}
\left(C_{2}^{LR}\right)_{\tilde g}
\simeq
-\frac{\alpha_s^2}{m_{\tilde q}^{2}}
\left[(\delta_d^{LL})_{ij}(\delta_d^{RR})_{ij}\right]
g_4^{(1)}(x_g)~,
\end{eqnarray}
where $(\delta_d^{LL})_{ij},(\delta_d^{RR})_{ij}$ are the mass insertion (MI) parameters
(as defined in ref.~\cite{Altmannshofer:2009ne}), $x_g=M_{\tilde g}^2/m_{\tilde q}^{2}$,
the loop function is such that $g_4^{(1)}(1)=23/180$ and $ij=23,13,12$ for the $B^{0}_s$,
$B^{0}_d$ and $K^0$ systems, respectively.

The dominant Higgs mediated contributions for the $B^{0}_s$ system read
\begin{eqnarray}
\left(C_{2}^{LR}\right)_{H}
\!\!&\simeq&\!\!
\frac{\alpha_2^2 \alpha_s}{4\pi}\frac{m_b^2 m_t^2}{2 M_W^4}
\frac{t^{4}_{\beta}}{(1+\epsilon t_{\beta})^{4}}
\frac{|\mu|^2}{M_A^2 m_{\tilde q}^{2}}
\nonumber\\
&\times&\!\!\!
\left[\frac{A_t^* M_{\tilde g}}{m_{\tilde q}^{2}}
h_1(x_g) h_3(x_\mu)\right](\delta^{RR}_d)_{23}V_{ts}~,
\label{eq:C4HB}
\end{eqnarray}
where $t_{\beta}=\tan\beta$, $x_\mu=|\mu|^2/m_{\tilde q}^{2}$, $h_1(1)=4/9$, $h_3(1)=-1/4$, $|\epsilon|\approx 10^{-2}$ is the well known resummation factor stemming from ($t_{\beta}$
enhanced) non-holomorphic threshold corrections and $m_b$ has to be evaluated at the scale
$M_A$. Notice that, the presence of the RR MIs is crucial to avoid the typical
$m_s/m_b$ suppression of $\left(C_{2}^{LR}\right)_{H}$ arising within a MFV scenario~\cite{Buras:2001mb}.
The Wilson coefficient for $B^{0}_d$ mixing can be obtained by replacing 23 with 13 and
$V_{ts}$ by $V_{td}$.

In the case of $K^0$ mixing, the most relevant effect from the neutral Higgses arises
only at the fourth order in the MI expansion. Since in the $SSU(5)_{RN}$
model the MI are radiatively induced, this effect can be safely neglected.

Moreover, we remind that in contrast to the case with gluino box contributions, there
are no analogous Higgs mediated contributions for $D^0-\bar D^0$ mixing as there are
no $\tan\beta$ enhanced non-holomorphic threshold corrections in the up-quark sector.

{\bf 2.} The CP asymmetries $S_f$ in the decays of neutral $B^{0}_d$ mesons into final
CP eigenstates $f$ can be affected by NP both in the $B^{0}_d$ mixing amplitude and 
the decay amplitudes $\bar b\to\bar qq\bar s$ ($q=s,d,u$).
In the $SSU(5)_{RN}$ model, only the latter effect can be sizable.
The asymmetries are defined as
\begin{equation}
S_f=\frac{2{\rm Im}(\lambda_f)}{1+|\lambda_f|^2}~.
\label{eq:Sf}
\end{equation}
where $\lambda_f=e^{-2i(\beta + \varphi_{B_d})}(\overline{A}_f/A_f)$ with $\varphi_{B_d}$
being the NP phase of the $B^{0}_d$ mixing amplitude, $M_{12}^d$, and $A_f$ ($\overline{A}_f$)
is the decay amplitude for $B^{0}_d(\overline{B^{0}_d})\to f$. Within the SM it turns out that
$A_f^{\rm SM}=A_f^c(1+a_f^u e^{i\gamma})$ where the $a_f^u$ parameters have been evaluated
in the QCD factorization approach at the leading order and to zeroth order in $\Lambda/m_b$ in~\cite{buchalla}.
In the presence of NP one can generally define the amplitude $A_f$ as
\begin{equation}
A_f = A_{f}^{\rm SM} + A_f^c\sum_i\left(b_{fi}^c+b_{fi}^ue^{i\gamma}\right)
\left(C_i^{*}+\zeta \tilde C_i^{*}\right)~,
\label{eq:def_b_fu}
\end{equation}
where $C_i$ and $\tilde{C}_i$ are the NP contributions to the Wilson coefficients evaluated
at the scale $M_W$ (see Ref.~\cite{Altmannshofer:2009ne} for the notation), the parameters
$b_{fi}^u$ and $b_{fi}^c$ calculated in~\cite{buchalla}
and $\zeta=\pm 1$ depending on the parity of the final state; for instance $\zeta=1$ for
$\phi K_S$ and $\zeta=-1$ for $\eta^\prime K_S$. Within the $SSU(5)_{RN}$ model the by far
dominant contribution to $A_f$ is provided by $\tilde C_8$ which reads~\cite{Altmannshofer:2009ne}
\begin{eqnarray}
\frac{4 G_F}{\sqrt{2}}~\tilde C_{8}^{\tilde g}
\simeq
\frac{g_s^2}{m_{\tilde q}^{2}}
\frac{M_{\tilde g}\mu^*}{m_{\tilde q}^{2}} \frac{t_{\beta}}{(1+\epsilon t_{\beta})}
\frac{(\delta_d^{RR})_{32}}{V_{ts}^*}g_{8}(x_g)~.
\label{eq:tildeC7_g}
\end{eqnarray}
where the loop function is such that $g_{8}(1)=-7/120$.

%%%%%%%%%%%%%%%%%%%%%%%%%%%           B_s -> mu+mu-         %%%%%%%%%%%%%%%%%%%%%%%%%%%%%%%

The branching ratio for $B_q\to\mu^+\mu^-$ (with $q=s,d$) in the presence of NP scalar
currents can be expressed as~\cite{Altmannshofer:2009ne}
\begin{equation}
\frac{{\rm BR}(B_q \!\to\! \mu^+ \mu^-)}{{\rm BR}(B_q \!\to\! \mu^+ \mu^-)_{\rm SM}}
\simeq \left|1 - C^q_S -\tilde C^q_S\right|^2 + \left|C^q_S - \tilde C^q_S\right|^2~,
\end{equation}
where the relevant contributions to the Wilson coefficients $C^q_S$ and $\tilde C^q_S$
arising in the $SSU(5)_{RN}$ model read

\begin{eqnarray}
\tilde C^q_S \!\!&\simeq&\!\! \frac{\alpha_2 \alpha_s}{8M^2_A} 
\frac{m^{2}_{B_q}}{M_W^2 C_{10}^{\rm SM}}
\frac{t^{3}_{\beta}}{(1\!+\!\epsilon t_{\beta})^{2}}
\frac{M_{\tilde g}\mu^*}{m_{\tilde q}^{2}} (\delta_d^{RR})_{3q} h_1(x_g)~,
\label{eq:CStilde}
\nonumber\\
C^q_S \!\!&\simeq&\!\! -\frac{\alpha_2^2}{8M^2_A} 
\frac{m^{2}_{B_q}}{M_W^4 C_{10}^{\rm SM}}
\frac{m_t^2 t^{3}_{\beta}}{(1\!+\!\epsilon t_{\beta})^{2}}
\frac{A_t\mu}{m_{\tilde q}^{2}}V_{tq}^* h_3(x_\mu)~,
\label{eq:CS}
\end{eqnarray}
with $C^{\rm SM}_{10}$ given for instance in~\cite{Altmannshofer:2009ne}.

%%%%%%%%%%%%%%%%%%%%%%%%%%%%%%%%         EDMs        %%%%%%%%%%%%%%%%%%%%%%%%%%%%%%%%%%%%%%

{\bf 3.} Also the hadronic and leptonic EDMs might be generated by {\it flavor dependent} phases
(flavored EDMs)~\cite{Hisano:2008hn}. In particular, the neutron EDM $d_n$ can be estimated
from the naive quark model as $d_n\approx\frac{4}{3} d_d-\frac{1}{3} d_u$ (with $d^{(c)}_{f}$
evaluated at $1$~GeV) or, alternatively, by means of QCD sum rule techniques~\cite{qcdsumrules,Demir:2002gg,Demir:2003js,Olive:2005ru}.
In the latter case, it turns out that
\beq
 d_n \!= (1\pm 0.5)\Big[ 1.4 (d_d-0.25 d_u) + 1.1 e\, (d^c_d+0.5 d^c_u)\Big]\,.
\label{Eq:dn_odd}
\eeq
Similarly, the prediction for the Mercury EDM in the QCD sum rule approach reads~\cite{Demir:2003js,Olive:2005ru}
\bea
\label{eq:mercuryedm}
d_{\rm Hg}
&\simeq&
7\times 10^{-3}\,e\,(d_u^c-d_d^c)\,,
\eea
where, in eq.~(\ref{eq:mercuryedm}), we have retained only the contributions relevant to
our analysis. Notice that, the values of $d^{(c)}_{f}$ entering the EDM's predictions of
eqs.~(\ref{Eq:dn_odd},\ref{eq:mercuryedm}) are assumed to be evaluated at $1$~GeV
by means of QCD renormalization group evolution~\cite{Degrassi:2005zd} from the corresponding
values at the electroweak scale.

The dominant gluino/squark contribution to the down-quark (C)EDMs at the SUSY scale reads
\beq
\left\{\frac{d_{d_i}}{e},~d^c_{d_i}\right\}\!\approx\!
\frac{\alpha_{s}}{4\pi}\frac{m_{b}}{m_{\tilde q}^{2}}
\frac{M_{\tilde g}\mu}{m_{\tilde q}^{2}} t_{\beta}
{\rm Im}\left[(\delta^{LL}_{d})_{i3}(\delta^{RR}_{d})_{3i}\right]
f(x_g)\,,
\label{Eq:edm_d_gluino}
\eeq
where the loop functions satisfy $f(1)=$ $\{4/135,~11/180\}$.
\begin{figure*}[th]
\includegraphics[width=0.4\textwidth]{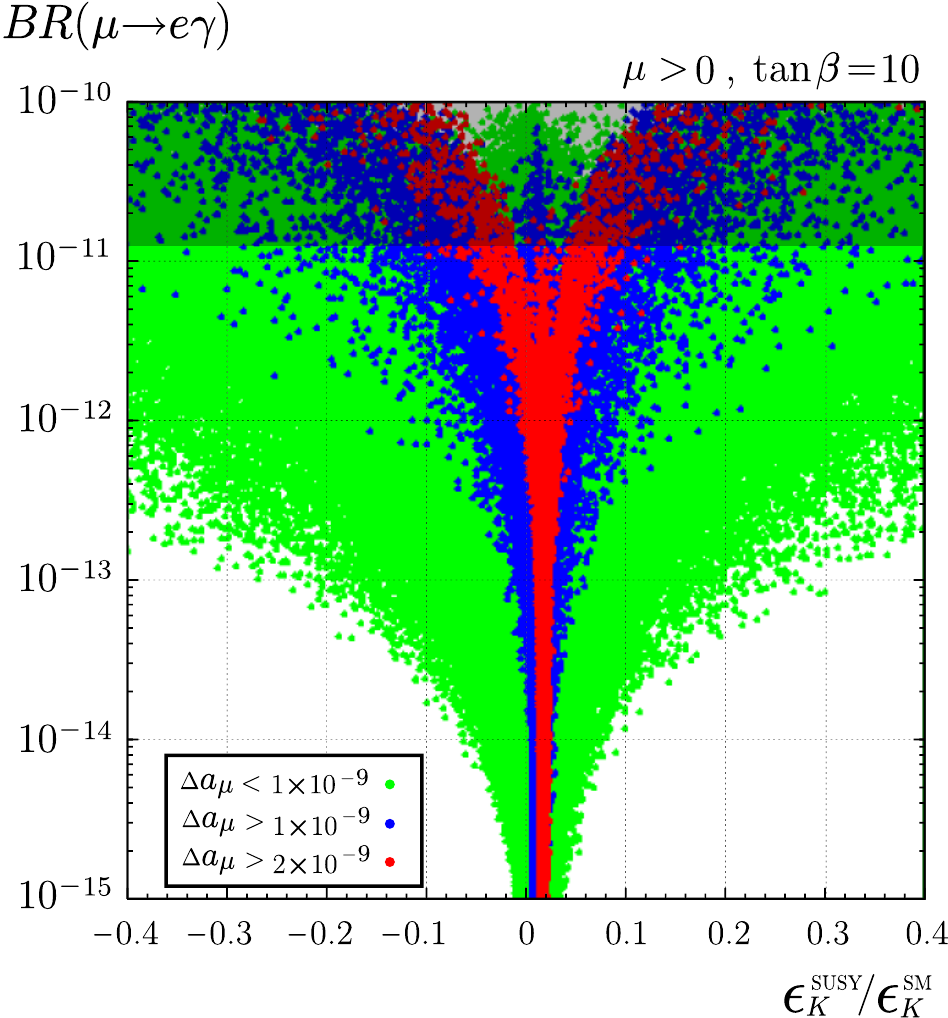}~
\includegraphics[width=0.4\textwidth]{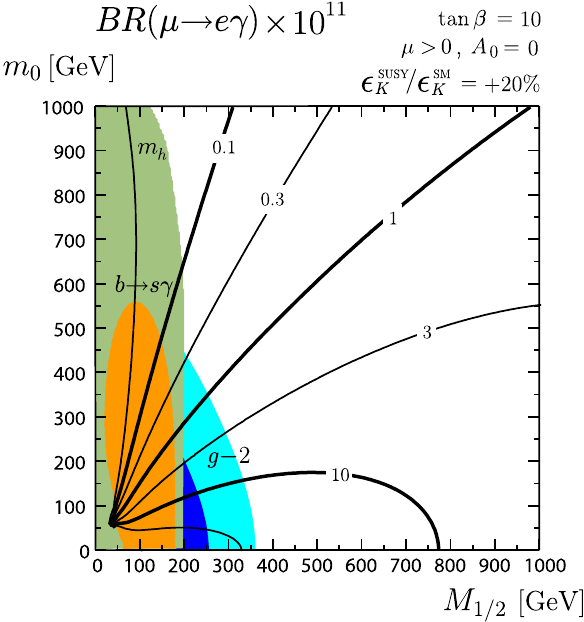}\\
\includegraphics[width=0.4\textwidth]{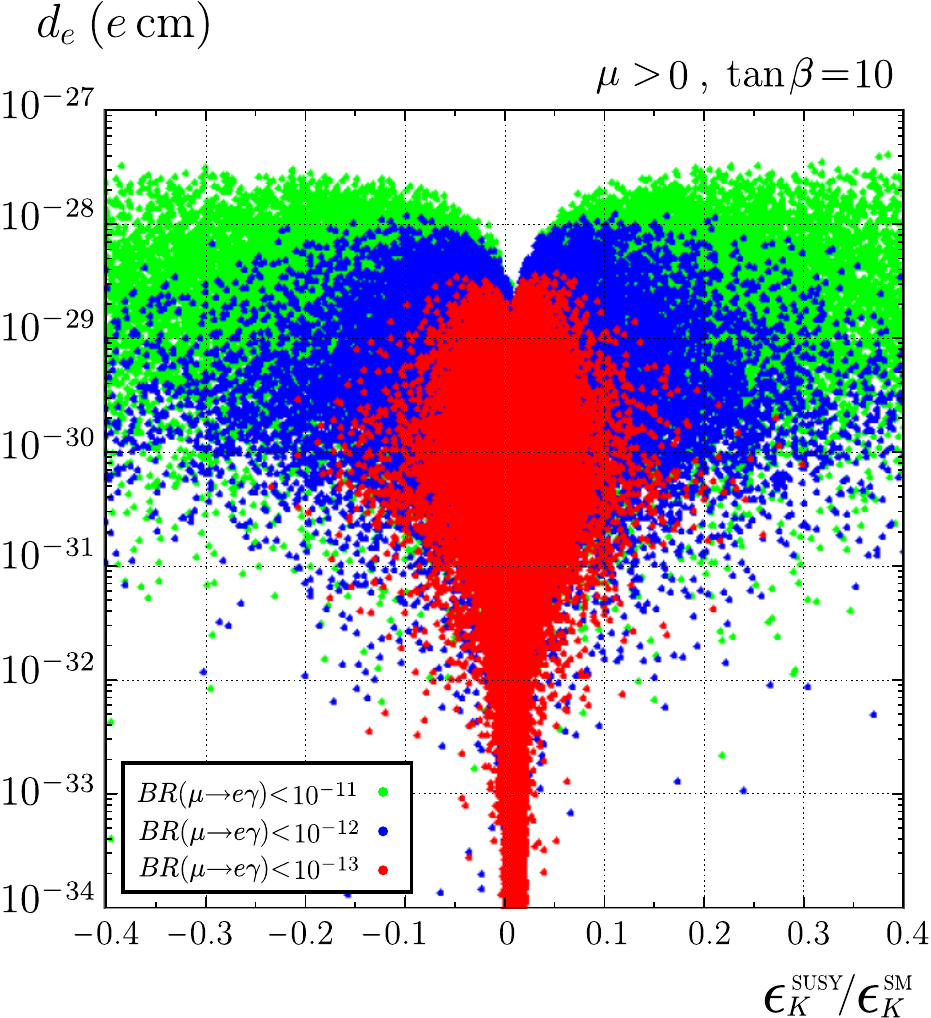}~
\includegraphics[width=0.4\textwidth]{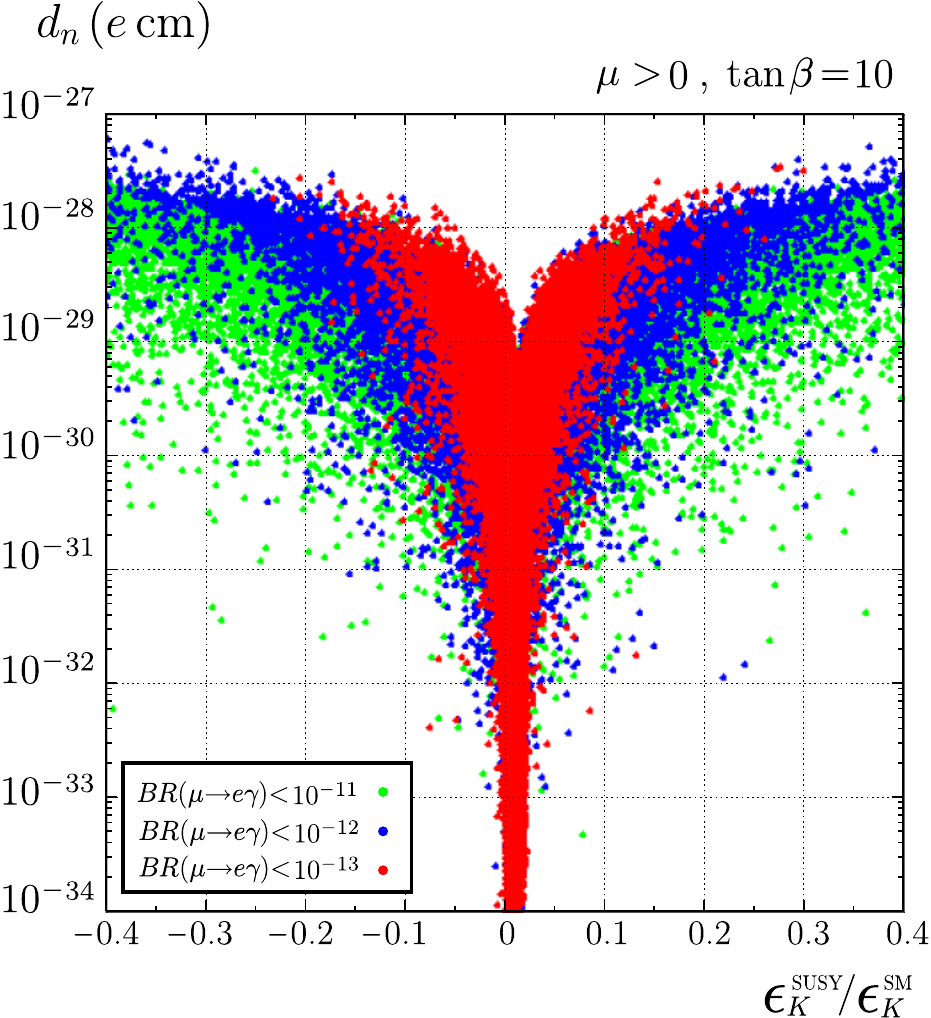}
\caption{
Upper left: ${\rm BR}(\mu\to e\gamma)$ vs. $\epsilon^{SUSY}_{K}/(\epsilon^{SM}_{K})$.
The blue (red) points can explain the $(g-2)_{\mu}$ anomaly at the level of
$\Delta a^{\rm SUSY}_{\mu}\gtrsim 1(2)\times 10^{-9}$.
Upper right: ${\rm BR}(\mu\to e\gamma)$ in the $(m_0,M_{1/2})$ plane imposing
a $+20\%$ NP effect in $\epsilon^{SUSY}_{K}$.
Lower left (right): electron (neutron) EDM vs. $\epsilon^{SUSY}_{K}/(\epsilon^{SM}_{K})$
for different values of ${\rm BR}(\mu\to e\gamma)$. In all the plots we vary the SUSY
parameters in the ranges $m_0<1\,{\rm TeV}$, $M_{1/2}<1\,{\rm TeV}$, $|A_0|<3m_0$
(in the only upper right plot we set $|A_0|=0$), $\tan\beta=10$ and $\mu>0$.
We assume a hierarchical spectrum for both light and heavy neutrinos setting
$m_{\nu_3}=0.05{\rm eV}$ and varying the neutrino parameters in the ranges
$10^{11}\leq M_{\nu_3}({\rm GeV})\leq 10^{15}$, $10^{-5}\leq U_{e3}\leq 0.1$.}
\label{fig2}
\end{figure*}
\begin{figure*}[th]
\includegraphics[width=0.4\textwidth]{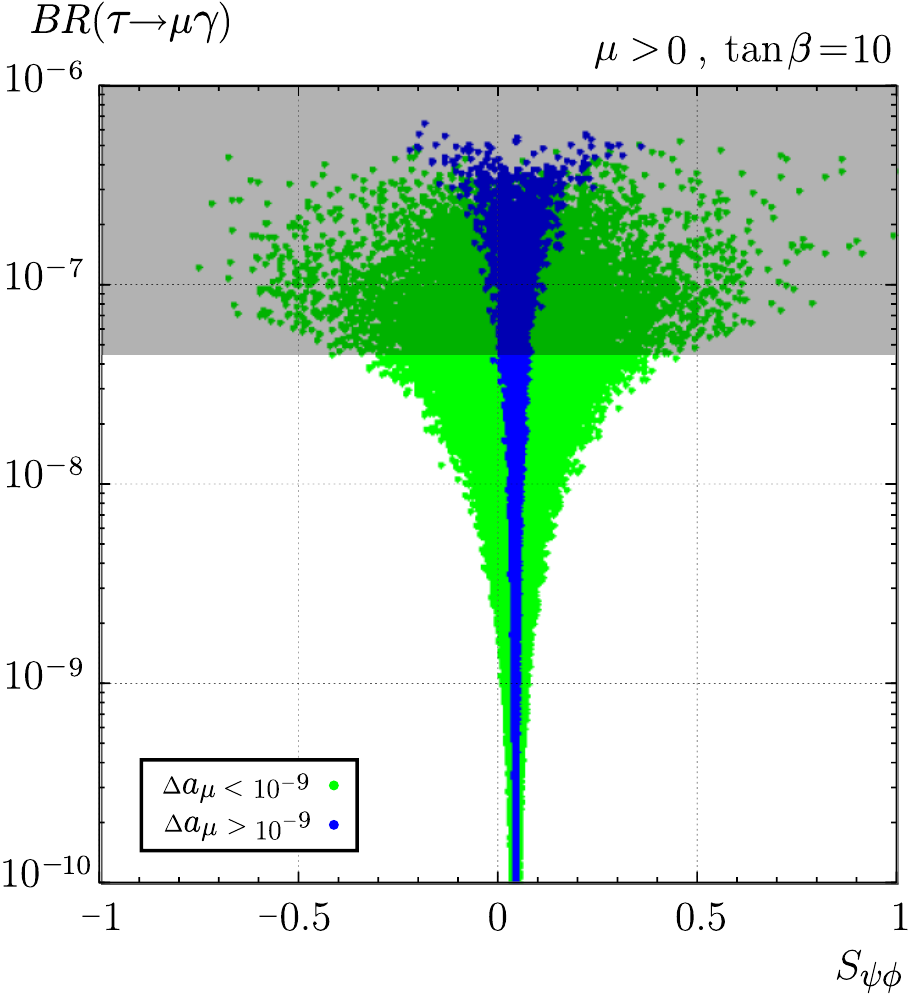}~
\includegraphics[width=0.4\textwidth]{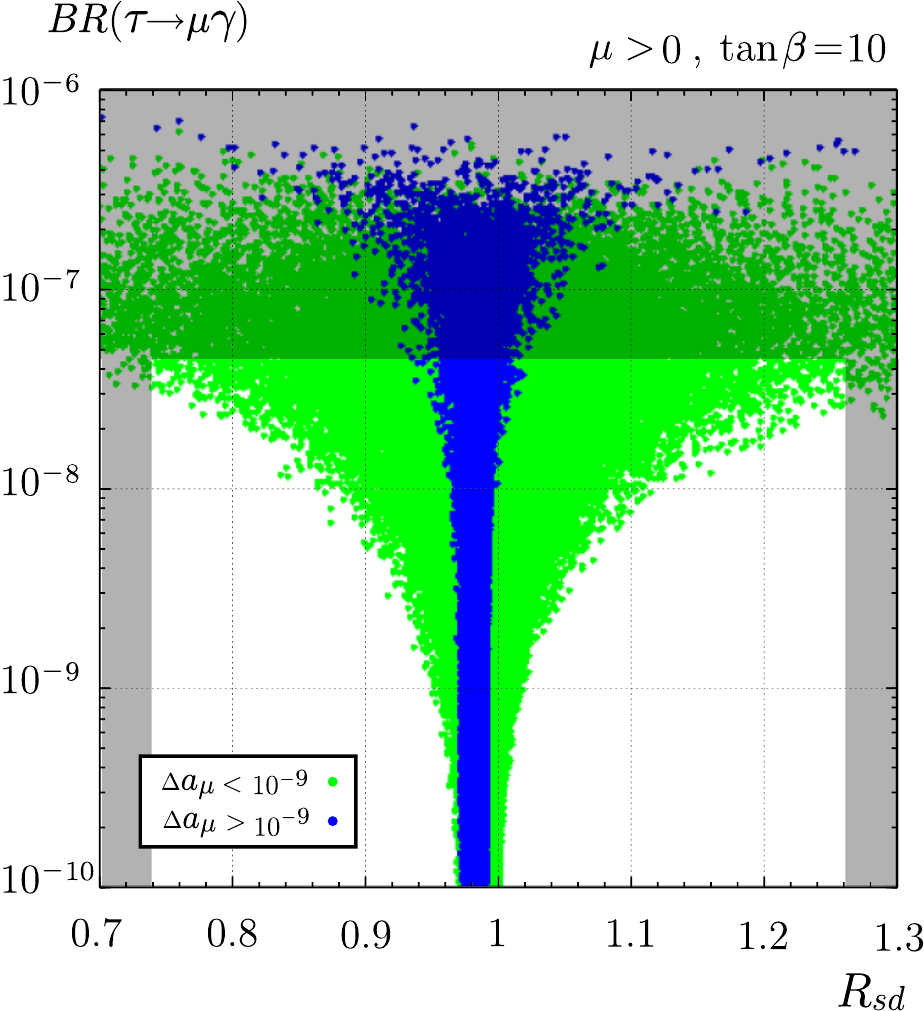}\\
\includegraphics[width=0.4\textwidth]{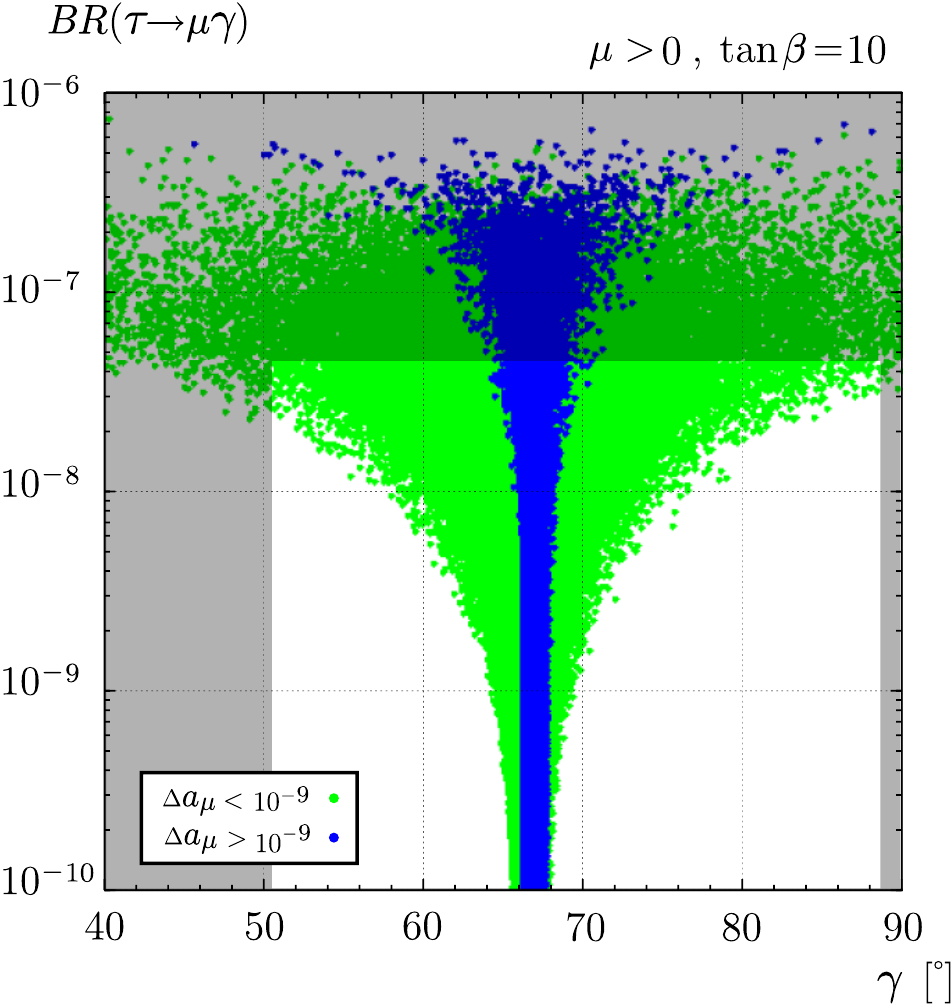}~
\includegraphics[width=0.4\textwidth]{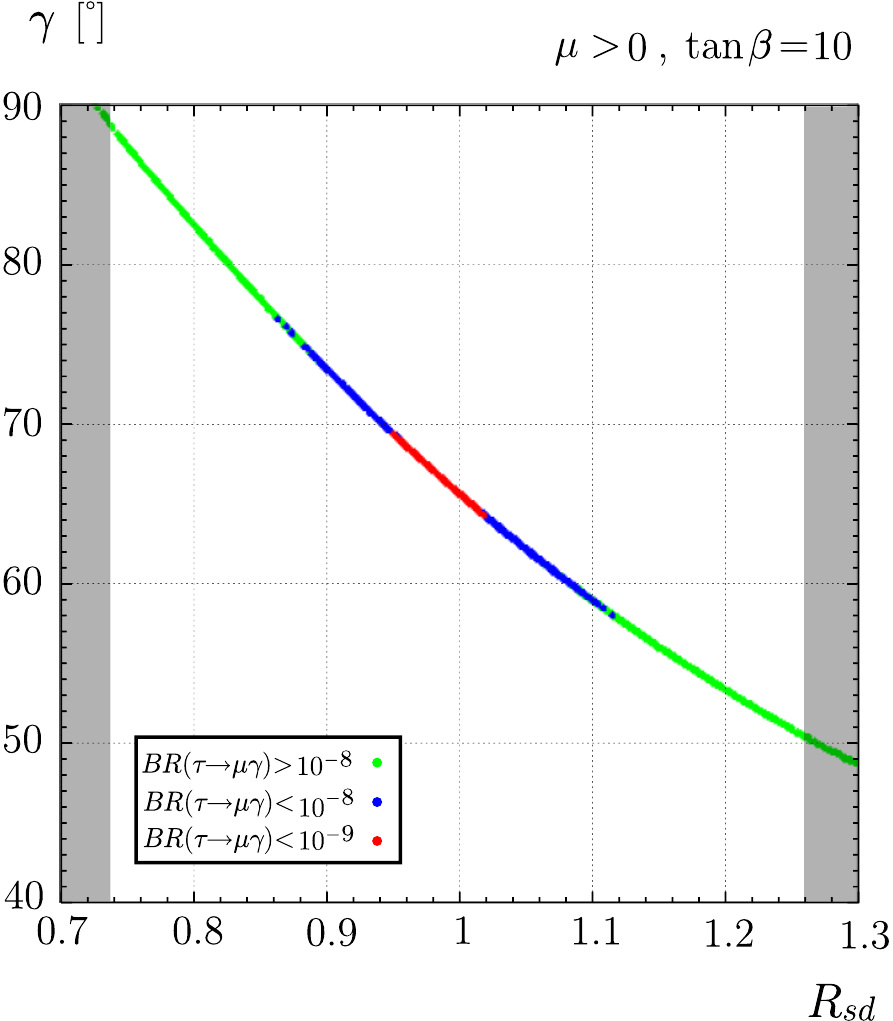}
\caption{
Upper left: ${\rm BR}(\tau\to\mu\gamma)$ vs. $S_{\psi\phi}$.
Upper right: ${\rm BR}(\tau\to\mu\gamma)$ vs. $R_{sd}$.
Lower left: ${\rm BR}(\tau\to\mu\gamma)$ vs. $\gamma$.
Lower right: $\gamma$ vs. $R_{sd}$.
Blue points satisfy the condition $\Delta a^{\rm SUSY}_{\mu}\gtrsim 1\times 10^{-9}$
while the grey bands indicate the experimentally excluded regions.
In all the plots we vary the SUSY parameters in the ranges $m_0<1\,{\rm TeV}$,
$M_{1/2}<1\,{\rm TeV}$, $|A_0|<3m_0$, $\tan\beta=10$ and $\mu>0$.
We assume a hierarchical spectrum for both light and heavy neutrinos setting
$m_{\nu_3}=0.05{\rm eV}$, $U_{e3}= 0$ and varying the heaviest heavy neutrino
mass in the ranges $10^{13}\leq M_{\nu_3}({\rm GeV})\leq 10^{15}$.}
\label{fig3}
\end{figure*}
\begin{figure*}[th]
\includegraphics[width=0.4\textwidth]{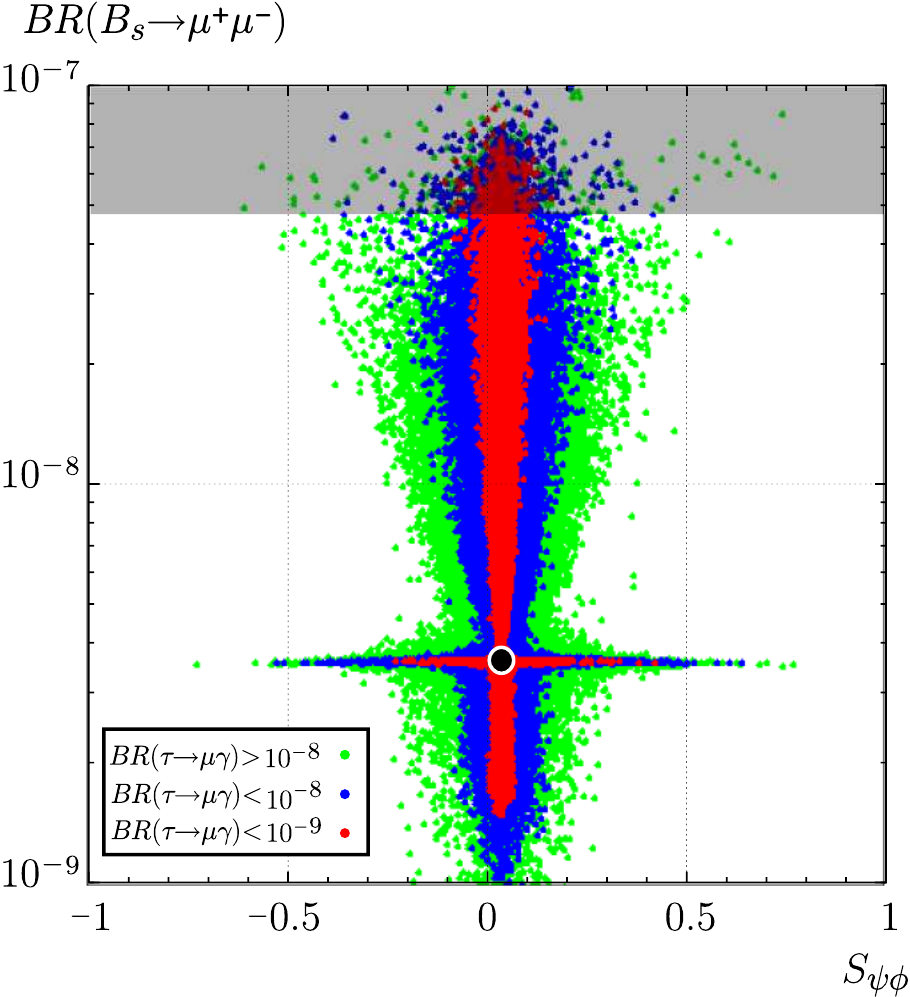}~
\includegraphics[width=0.4\textwidth]{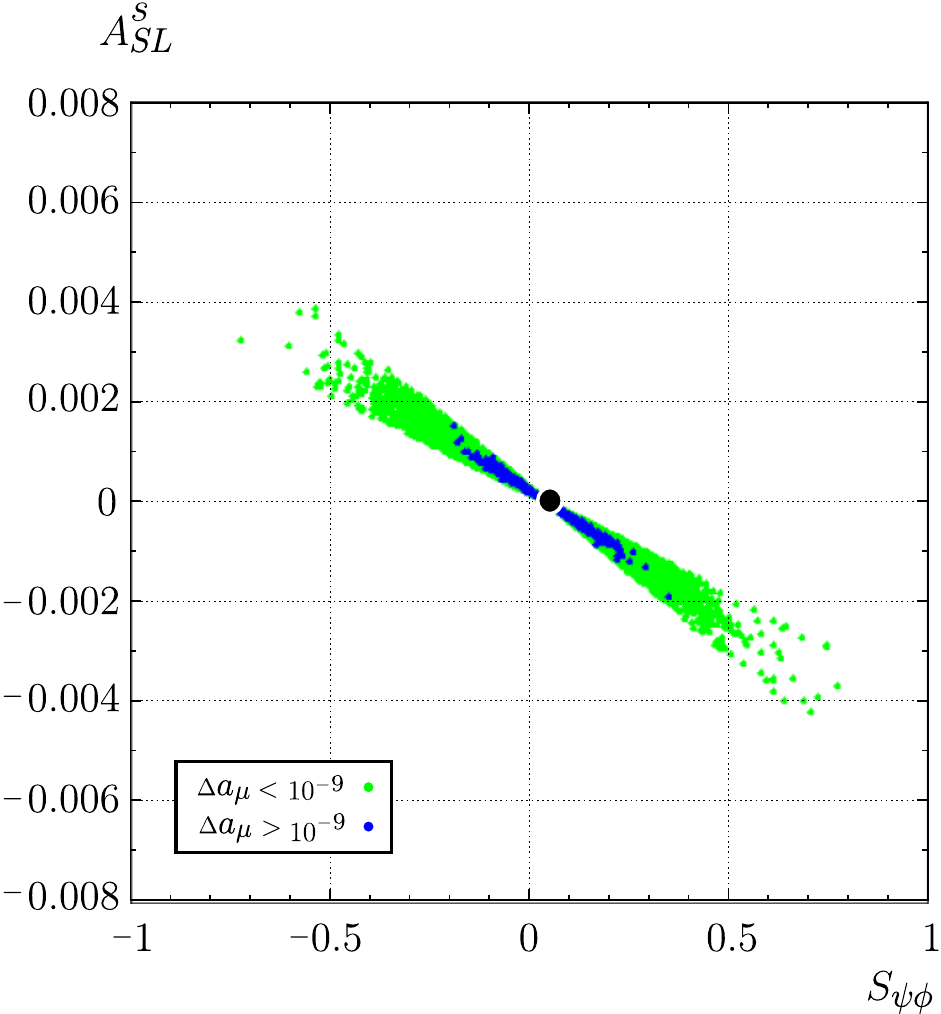}\\
\includegraphics[width=0.4\textwidth]{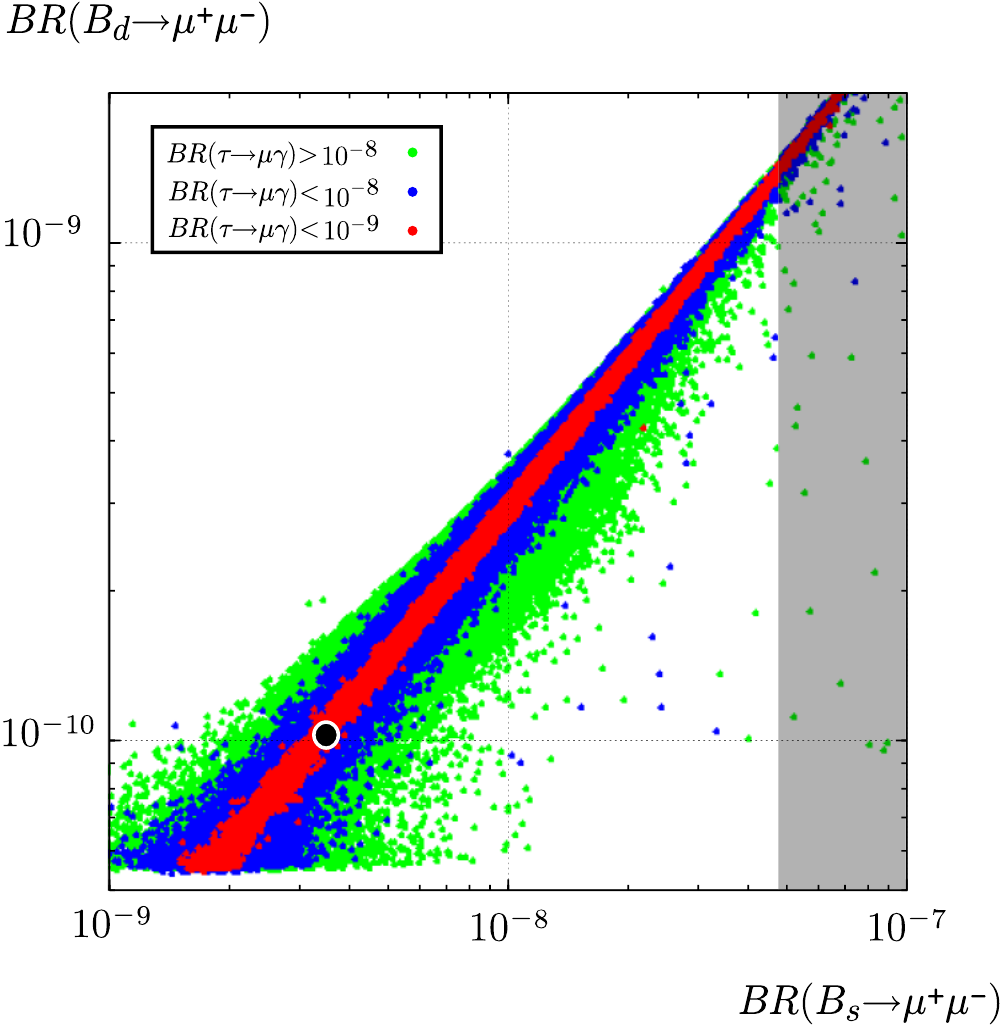}~
\includegraphics[width=0.4\textwidth]{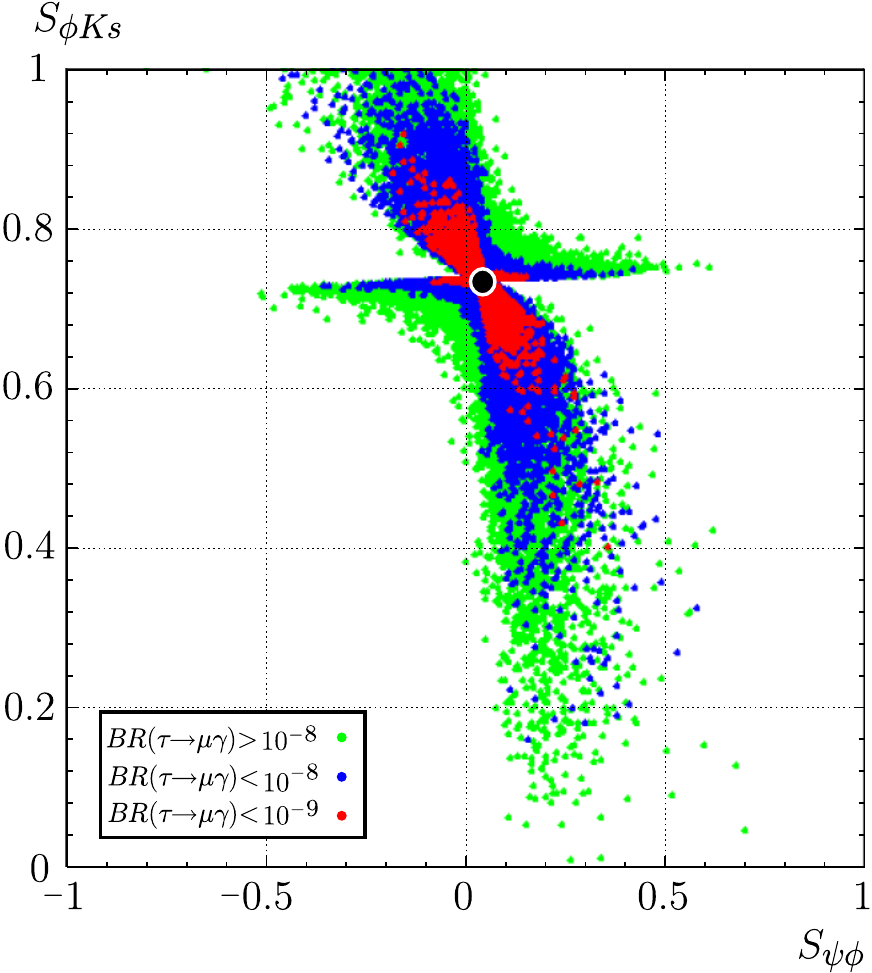}\\
\includegraphics[width=0.4\textwidth]{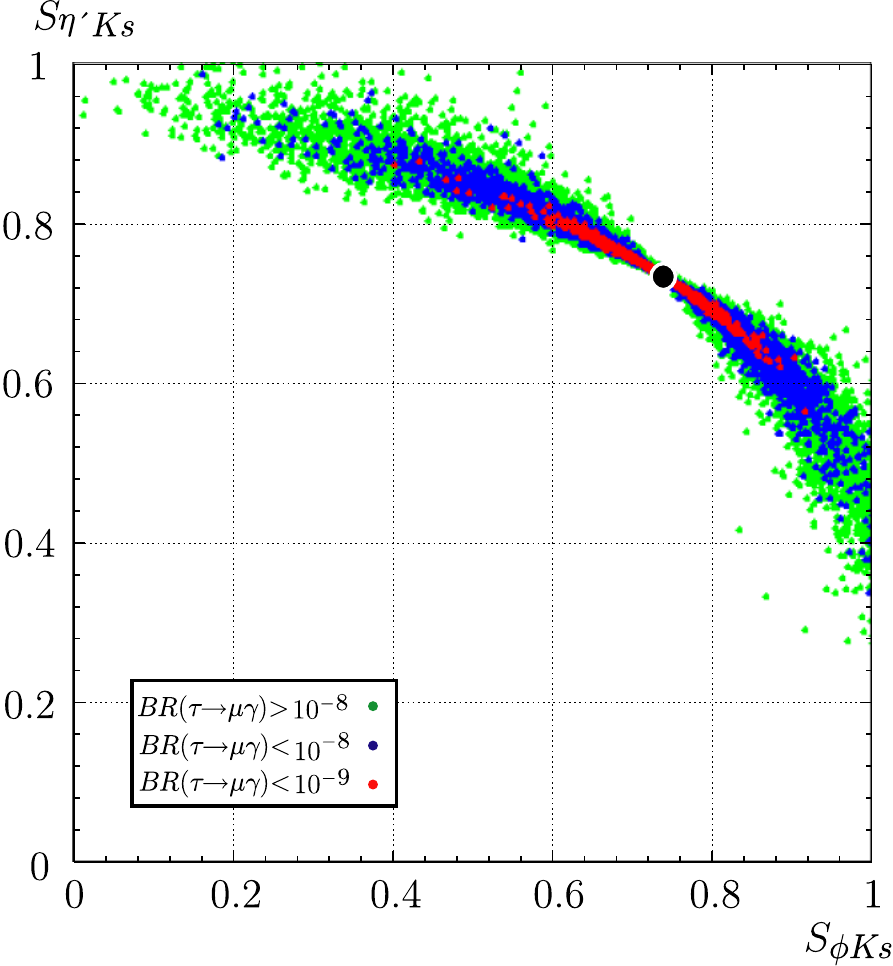}~
\includegraphics[width=0.4\textwidth]{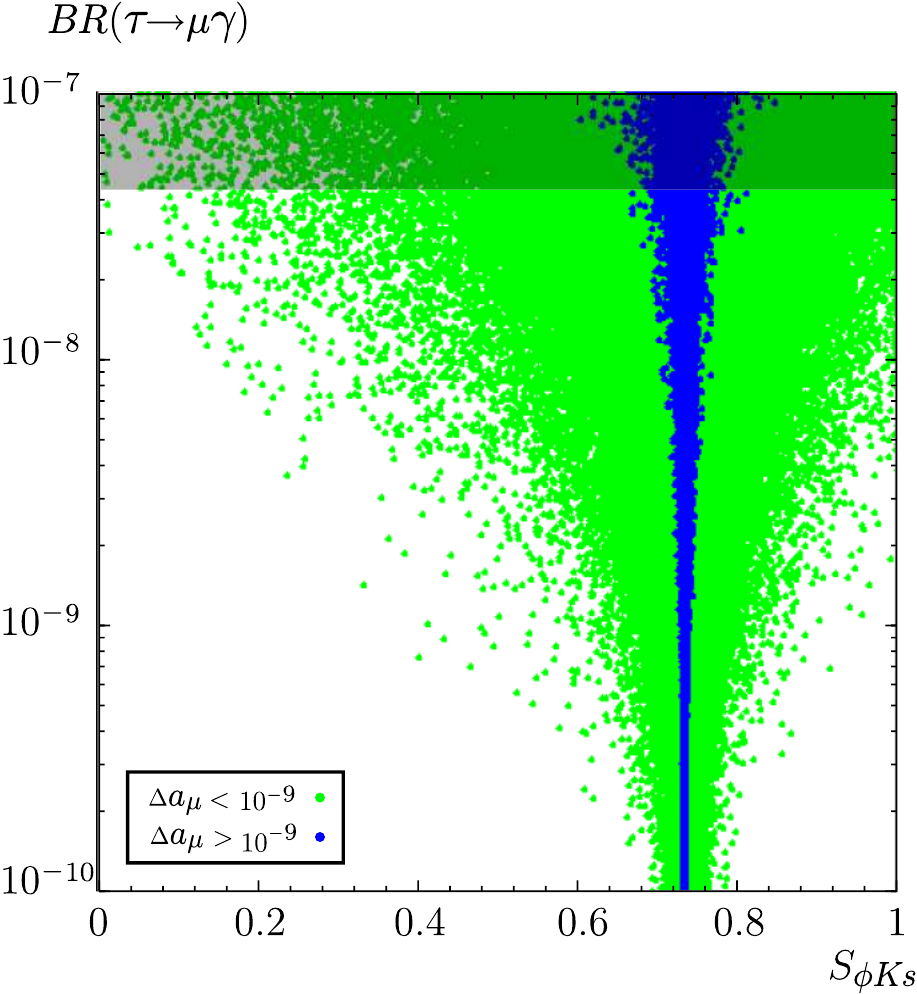}
\caption{
Upper left: ${\rm BR}(B_s\to \mu^+\mu^-)$ vs. $S_{\psi\phi}$.
Upper right: $S_{\psi\phi}$ vs. $A^{s}_{SL}$.
Central left: ${\rm BR}(B_d\to \mu^+\mu^-)$ vs. ${\rm BR}(B_s\to \mu^+\mu^-)$.
Central right: $S_{\psi\phi}$ vs. $S_{\phi K_S}$.
Lower left: $S_{\eta^{\prime} K_S}$ vs. $S_{\phi K_S}$.
Lower right: ${\rm BR}(\tau\to\mu\gamma)$ vs. $S_{\phi K_S}$.
The grey bands indicate the experimentally excluded regions.
In all the plots we vary the SUSY parameters in the ranges $m_0<2\,{\rm TeV}$,
$M_{1/2}<1\,{\rm TeV}$, $|A_0|<3m_0$, $\tan\beta\leq 60$ and $\mu>0$.
We assume a hierarchical spectrum for both light and heavy neutrinos setting
$m_{\nu_3}=0.05{\rm eV}$, $U_{e3}= 0$ and varying the heaviest heavy neutrino
mass in the ranges $10^{13}\leq M_{\nu_3}({\rm GeV})\leq 10^{15}$.}
\label{fig4}
\end{figure*}

%%%%%%%%%%%%%%%%%%%%%%%%%%%%%%%%%%%%%%%%%%%%%%%%%%%%%%%%%%%%%%%%%%%%%%%%%%%%%%%%%%%%%%%%%%%%%%
\section{Leptonic sector}
%%%%%%%%%%%%%%%%%%%%%%%%%%%%%%%%%%%%%%%%%%%%%%%%%%%%%%%%%%%%%%%%%%%%%%%%%%%%%%%%%%%%%%%%%%%%%%

The branching ratio for $\ell_{i}\rightarrow \ell_{j}\gamma$ can be written as
\bea
\frac{{\rm BR}(\ell_{i}\rightarrow  \ell_{j}\gamma)}{{\rm BR}(\ell_{i}\rightarrow 
\ell_{j}\nu_i\bar{\nu_j})} =\frac{48\pi^{3}\alpha}{G_{F}^{2}}(|A_L^{ij}|^2+|A_R^{ij}|^2)\,.
\nonumber
\eea
Starting from the full expressions of ref.~\cite{Hisano:1995cp} (which we use in our numerical analysis), and specializing to the illustrative case of a degenerate SUSY spectrum with a common
mass $m_{\tilde\ell}$, one can find
\beq
  A^{ij}_L \simeq
  \frac{\alpha_2}{60\pi}\frac{t_{\beta}}{m^{2}_{\tilde\ell}}
  (\delta^{LL}_{\ell})_{ji}~.
\label{MI_degenerate}
\eeq
The main SUSY contribution to $a^{\rm MSSM}_\mu$ is usually provided by the loop
exchange of charginos and sneutrinos. In the limit of degenerate SUSY masses one 
finds
\beq
\frac{a^{\rm MSSM}_\mu}{ 1 \times 10^{-9}}
\approx 1.5\left(\frac{t_{\beta}}{10} \right)
\left( \frac{300~\rm GeV}{m_{\tilde \ell}}\right)^2 \text{sgn}\,\mu\,.
\label{eq:g_2}
\eeq
Assuming a degenerate SUSY spectrum, it is straightforward to find the
correlation between $\Delta a^{\rm SUSY}_{\mu}$ and the branching ratios 
for $\ell_i\to\ell_j\gamma$~\cite{Hisano:2001qz,Hisano:2009ae}
\begin{eqnarray}
{\rm BR}(\mu\to e\gamma)&\approx&
2\times 10^{-12}
\left[\frac{\Delta a^{\rm SUSY}_{\mu}}{ 3 \times 10^{-9}}\right]^{2}
\bigg|\frac{(\delta^{LL}_{\ell})_{21}}{10^{-4}}\bigg|^2\,,\nonumber\\
{\rm BR}(\tau\to\mu\gamma)&\approx&
8\times 10^{-8}
\left[\frac{\Delta a^{\rm SUSY}_{\mu}}{ 3 \times 10^{-9}}\right]^{2}
\bigg|\frac{(\delta^{LL}_{\ell})_{32}}{10^{-2}}\bigg|^2\,,
\label{lfvgm2}
\end{eqnarray}
where we have assumed that the MIs $(\delta^{LL}_{\ell})_{ij}$ provide the
dominant contributions to ${\rm BR}(\ell_i\to \ell_j\gamma)$, as it happens in the GUT 
framework analyzed here.

Within the $SSU(5)_{RN}$ model, leptonic EDMs are generated via {\it flavour dependent}
phases (flavoured EDMs). It turns out that
\beq
\frac{d_{\ell_i}}{e}\simeq
-\frac{\alpha_Y}{4\pi}\bigg(\frac{m_\tau}{m_{\tilde \ell}^{2}}\bigg)\,t_{\beta}~
\frac{{\rm Im}[\left(\delta^{RR}_{\ell}\right)_{i3}\left(\delta^{LL}_{\ell}\right)_{3i}]}{30}\,,
\label{edm_flavor}
\eeq
where a common SUSY mass $m_{\tilde \ell}$ has been assumed. If $t_{\beta}=10$ and
$m_{\tilde \ell}=300\,{\rm GeV}$, it turns out that $d_{\ell_i}\sim 10^{-22}\times
{\rm Im}[(\delta^{RR}_{\ell})_{i3}(\delta^{LL}_{\ell})_{3i}]\,e\,$cm.

\section{Hadron-lepton correlations}
\label{correlations}

As already anticipated in the Introduction, SUSY GUT models link flavor-violating observables
of the leptonic and hadronic sectors. In the following, we provide approximate analytical
expressions for these hadron-lepton correlations in order to get an idea of where we stand.
In particular, the GUT relation $(\delta_e^{LL})_{ij}=(\delta_d^{RR})_{ji}$, which is modified
at the electroweak scale as $(\delta_e^{LL})_{ij}=(\delta_d^{RR})_{ji} \times (m_{\tilde{q}}/m_{\tilde{\ell}})^2$, implies
\bea
{\rm BR}(\mu\to e\gamma)
&\approx&
2\times 10^{-12}
\left(\frac{\epsilon^{SUSY}_{K}}{10^{-4}}\right)^{2}
\left(\frac{t_\beta}{10}\right)^2
\nonumber\\
&\times&
\bigg|\frac{(\delta_d^{LL})_{12}}{10^{-4}}\bigg|^{-2}
\left(\frac{m_{\tilde{q}}}{m_{\tilde{\ell}}}\right)^8\,,
\label{mueg_epsk}
\eea
and similarly for the 23 sector
\bea
{\rm BR}(\tau\to\mu\gamma)&\approx&
7\times 10^{-8}
\bigg[ (C_{B_s}-1)^2 + C_{B_s} S_{\psi\phi}^2 \bigg]
\nonumber\\
&\times&
\left(\frac{t_\beta}{10}\right)^2
\bigg|\frac{(\delta_d^{LL})_{23}}{10^{-2}}\bigg|^{-2}
\left(\frac{m_{\tilde{q}}}{m_{\tilde{\ell}}}\right)^8\,,
\label{tmug_spsiphi}
\eea
where we have assumed a common SUSY mass $m_{\tilde{q}}$ ($m_{\tilde{\ell}}$) for the hadron
(lepton) sector and $(\delta_d^{LL})_{12,23}$ have been normalized to the typical values they
attain when they are radiatively generated by the large top Yukawa coupling and the CKM matrix
(see Eq.\,(\ref{Eq:SU5RN_FV})).

Finally, there is an even more direct correlation between ${\rm BR}(\tau\to\mu\gamma)$
and the NP effects entering $S_{\phi K_S}$.
Defining $S_{\phi K_S}=S_{\psi K_S}+\Delta S_{\phi K_S}$, we find
\bea
{\rm BR}(\tau\to\mu\gamma)&\approx&
3\times 10^{-8}\left(\frac{m_{\tilde{q}}}{m_{\tilde{\ell}}}\right)^8
\left|\Delta S_{\phi K_S}\right|^2\,,
\label{tmug_sphiks}
\eea
where, starting from eqs.~(\ref{eq:Sf}),(\ref{eq:def_b_fu}),(\ref{eq:tildeC7_g})
and keeping only linear terms in the NP contributions, it turns out that
$\Delta S_{\phi K_S}\approx -2 b_{\phi K_S}^c \cos 2\beta ~{\rm Im}\tilde{C_8}$
with $b_{\phi K_S}^c\approx 1.4$~\cite{buchalla}.

\section{Numerical analysis}
\label{sec:num_analysis}

In this section, we present the numerical results for the observables discussed in the
previous sections in the context of the $SSU(5)_{RN}$ model, assuming a gravity
mediated mechanism for the SUSY breaking terms with $M_P = 2.4\times 10^{18}$~GeV.

In the upper (left) plot of Fig.~\ref{fig2}, we show the predictions for ${\rm BR}(\mu\to e\gamma)$ vs. $\epsilon^{SUSY}_{K}/(\epsilon^{SM}_{K})$ varying the SUSY parameters in the
ranges $(m_0,M_{1/2})<1\,{\rm TeV}$, $|A_0|<3m_0$, $\tan\beta=10$ and $\mu>0$.
Concerning the neutrino sector, hereafter, we assume a hierarchical spectrum for both light
and heavy neutrinos such that $m_{\nu_3}=0.05{\rm eV}$ and we vary the neutrino parameters
in the ranges $10^{11}\leq M_{\nu_3}({\rm GeV})\leq 10^{15}$, $10^{-5}\leq U_{e3}\leq 0.1$.

The blue (red) points can explain the $(g-2)_{\mu}$ anomaly at the level of $\Delta a^{\rm SUSY}_{\mu}\gtrsim 1(2)\times 10^{-9}$ while satisfying the constraints from ${\rm BR}(B\to X_s\gamma)$~\cite{Amsler:2008zzb} at the $99\%$ C.L.. As we can see, sizable SUSY
effects in $\epsilon_{K}$, that might be desirable to solve the UT anomaly, generally
imply a lower bound for ${\rm BR}(\mu\to e\gamma)$ in the reach of the MEG experiment.
The above statement is even more strengthened if we further require to explain the
$(g-2)_{\mu}$ anomaly.

In the upper (right) plot of Fig.~\ref{fig2}, we show the values reached by
${\rm BR}(\mu\to e\gamma)$ in the $(m_0,M_{1/2})$ plane setting $\mu>0$, $A_0=0$,
$\tan\beta=10$ and imposing a NP effect in $\epsilon^{SUSY}_{K}$ at the level
of $+20\%$ compared to the SM contribution to solve the UT tension.
The grey region is excluded by the constraint from the lower bound on the lightest Higgs
boson mass $m_{h^0}$ (we impose $m_{h^0}>111.4$~GeV to take into account the theoretical 
uncertainties in the evaluation of $m_{h^0}$), the orange region is excluded by the
constraints on ${\rm BR}(B\to X_s\gamma)$ at the $99\%$ C.L. (we have evaluated
${\rm BR}(B\to X_s\gamma)$ including the SM effects at the NNLO~\cite{misiak} and the
NP contributions at the LO), the light blue (blue) region satisfies 
$\Delta a^{\rm SUSY}_{\mu}\gtrsim 1(2)\times 10^{-9}$.

In the lower plots of Fig.~\ref{fig2}, on the left (right), we show the electron (neutron)
EDM vs. $\epsilon^{SUSY}_{K}/(\epsilon^{SM}_{K})$ for different values of ${\rm BR}(\mu\to e\gamma)$.
The requirement of sizable non-standard effects in $\epsilon^{SUSY}_{K}$ always implies large
values for $d_{e,n}$, in the reach of the planned experimental resolutions, as well as values
for ${\rm BR}(\mu\to e\gamma)$ that are most likely within the MEG reach.
The correlations between leptonic and hadronic observables in the plots on the left
in Fig.~\ref{fig2} demonstrate very clearly that we deal here with a GUT scenario.

In figs.~\ref{fig3},~\ref{fig4}, we present the predictions for B-physics observables.
As discussed in the previous sections, within a $SSU(5)_{RN}$ model, $b\to s$
and $\tau\to\mu$ transitions are linked, therefore, processes like $B^{0}_s$ mixing
and $\tau\to\mu\gamma$ turn out to be related.

In the plots of fig.~\ref{fig3} we vary the SUSY parameters in the ranges 
$(m_0,M_{1/2})<1\,{\rm TeV}$, $|A_0|<3m_0$, $\tan\beta=10$ and $\mu>0$.
We assume a hierarchical spectrum for both light and heavy neutrinos setting
$m_{\nu_3}=0.05{\rm eV}$, $U_{e3}= 0$ and varying the heaviest heavy neutrino
mass in the range $10^{13}\leq M_{\nu_3}({\rm GeV})\leq 10^{15}$.

In the upper plot of fig.~\ref{fig3} on the left, we show the correlation between
${\rm BR}(\tau\to\mu\gamma)$ and $S_{\psi\phi}$. We see that, non-standard values for
$S_{\psi\phi}$ imply a lower bound for ${\rm BR}(\tau\to\mu\gamma)$ within the SuperB
reach. However, it seems unlikely to simultaneously explain the $(g-2)_{\mu}$ anomaly
(blue points correspond to $\Delta a^{\rm SUSY}_{\mu}\gtrsim 1\times 10^{-9}$) while
generating a large $S_{\psi\phi}$.
The situation can slightly change for large values of $\tan\beta$. In this case,
the constraints from ${\rm BR}(\tau\to\mu\gamma)$ might be still compatible with
$|S_{\psi\phi}|\leq 0.2$ and $\Delta a^{\rm SUSY}_{\mu}\gtrsim 1\times 10^{-9}$
(see fig.~\ref{fig4}).

In the upper plot of fig.~\ref{fig3} on the right, we show the correlation between
${\rm BR}(\tau\to\mu\gamma)$ vs. $R_{sd}$ defined as
\beq
 R_{sd}=\frac{\Delta M_d/\Delta M_s}{(\Delta M_d/\Delta M_s)_{\rm SM}}\,.
\eeq
We see that non-standard effects in $R_{sd}$ may be easily generated, providing a
possible solution to the UT anomaly. This will imply in turn a lower bound for
${\rm BR}(\tau\to\mu\gamma)$ within the expected experimental reach of a SuperB.

In the lower plot of fig.~\ref{fig3} on the left, we show the correlation between
${\rm BR}(\tau\to\mu\gamma)$ and the CKM angle $\gamma$. This correlation can 
be understood looking at the explicit expressions for $\gamma$ and $R_t$ in the presence
of NP, see eqs.(\ref{eq:Rb_gamma},\ref{eq:Rt_sin2beta_NP}). In particular,
NP effects in $R_{sd}$ would affect $R_t$ and therefore the determination of $\gamma$.
Moreover, since $b\to s$ and $\tau\to\mu$ transitions are linked in our framework,
it turns out that non-standard values for $\gamma$ imply a lower bound for
${\rm BR}(\tau\to\mu\gamma)$.
It will be exciting to monitor such a correlated NP effect at the LHCb.

In the lower plot of fig.~\ref{fig3} on the right, we report the correlation between
$\gamma$ vs. $R_{sd}$ clearly showing that negative NP effects in $R_{sd}$, accounting
for the UT anomaly, would imply large non-standard values for the angle $\gamma$.
Such large non-standard effects would also imply large (visible) values for
${\rm BR}(\tau\to\mu\gamma)$ as seen in the plot on the left.

In fig.~\ref{fig4}, we show the predictions for B-physics and lepton observables
including the large $\tan\beta$ regime, in order to make Higgs mediated effects for
$\Delta F = 1,2$ processes like ${\rm BR}(B_{s,d}\to\mu^+\mu^-)$ and $S_{\psi\phi}$
visible.

In particular, the plots of fig.~\ref{fig4} have been obtained employing the following
scan over the SUSY parameters: $m_0 (M_{1/2})<2(1)\,{\rm TeV}$, $|A_0|<3m_0$,
$\tan\beta<60$ and $\mu>0$. We assume a hierarchical spectrum for both light and heavy
neutrinos setting $m_{\nu_3}=0.05{\rm eV}$, $U_{e3}= 0$ and varying the heaviest heavy
neutrino mass in the range $10^{13}\leq M_{\nu_3}({\rm GeV})\leq 10^{15}$.

In the upper plot of fig.~\ref{fig4} on the left, we show ${\rm BR}(B_s\to \mu^+\mu^-)$
vs. $S_{\psi\phi}$ visualizing also the values attained by ${\rm BR}(\tau\to\mu\gamma)$
with different colours.
As we can see, $S_{\psi\phi}$ can depart from the SM expectations irrespective of whether
${\rm BR}(B_s\to \mu^+\mu^-)$ is SM-like or not. The reason is that $S_{\psi\phi}$ receives
large effects from both gluino/squark box contributions and from $\tan\beta$ enhanced double penguin Higgs contributions. In contrast, in the case of ${\rm BR}(B_s\to \mu^+\mu^-)$ only
the latter contribution can be effective.

In the upper plot of fig.~\ref{fig4} on the right, we show the (almost) model-independent
correlation between $A^{s}_{SL}$ and $S_{\psi\phi}$ (see eq.~(\ref{eq:Ab_SL_Y})).
Green points fulfill all the current available constraints, while blue points further
explain the $(g-2)_{\mu}$ anomaly at the level of $\Delta a^{\rm SUSY}_{\mu}\gtrsim
1\times 10^{-9}$.
While large departures from the SM expectations for $A^{s}_{SL}$ are still allowed,
the large value reported by the Tevatron~\cite{Abazov:2010hv} (see eq.~(\ref{ASL_exp}))
cannot be accounted for within the $SSU(5)_{RN}$ model.

In the central plot of fig.~\ref{fig4} on the left, we show the correlation between
${\rm BR}(B_s\to\mu^+\mu^-)$ and ${\rm BR}(B_d\to\mu^+\mu^-)$. Interestingly enough,
we notice that sizable departures from the MFV predictions $|V_{ts}/V_{td}|^2$ imply
large values for ${\rm BR}(\tau\to\mu\gamma)$, well within the SuperB reach.

In the central plot of fig.~\ref{fig4} on the right, we show the predictions for
$S_{\psi\phi}$ and $S_{\phi K_S}$. Noteworthy enough, these observables can sizably
depart from the SM expectations in a correlated manner.
The pattern of correlation is twofold: for moderate/low $\tan\beta$ values
(corresponding to the almost vertical band), $S_{\psi\phi}$ receives the dominant
contributions from gluino/squark boxes and the correlation might be in agreement
with the current non-standard experimental data while for large $\tan\beta$ values
(corresponding to the almost horizontal band), $S_{\psi\phi}$ receives the dominant
contributions from double penguin Higgs exchanges and the correlation is opposite.
In any case, large non-standard effects for $S_{\psi\phi}$ and/or $S_{\phi K_S}$
always imply experimentally visible values for ${\rm BR}(\tau\to\mu\gamma)$.

In the lower plot of fig.~\ref{fig4} on the left, we report the correlation between
$S_{\eta^{\prime} K_S}$ and $S_{\phi K_S}$ clearly showing that these observables
exhibit opposite deviations with respect to the SM expectations.

This is understood remembering that the NP amplitudes for these processes can be written
as $A_{\rm NP}\sim C_i+\zeta \tilde C_i$ where $C_i$ and $\tilde C_i$ are the NP Wilson
coefficients and $\zeta=\pm 1$ depending on the parity of the final state which is $\zeta=1$
for $\phi K_S$ and $\zeta=-1$ for $\eta^\prime K_S$ (see Section~4).
Since in the $SSU(5)_{RN}$ model $\tilde C_i$ provide the largely dominant
effects, $S_{\eta^{\prime} K_S}$ and $S_{\phi K_S}$ are expected to show opposite departures
from the SM predictions as confirmed numerically. This is in contrast to scenarios like
the flavour-blind MSSM~\cite{Altmannshofer:2008hc}, the MSSM with MFV or models with purely left-handed currents where $C_i$ are dominant~\cite{Altmannshofer:2009ne}.

In the lower plot of fig.~\ref{fig4} on the right, we also show the correlation
between ${\rm BR}(\tau\to\mu\gamma)$ and $S_{\phi K_S}$ confirming that sizable NP
effects for $S_{\phi K_S}$ imply a lower bound for ${\rm BR}(\tau\to\mu\gamma)$ within
the SuperB reach. However, we notice that an explanation of the $(g-2)_{\mu}$ anomaly
would prevent large non-standard effects for $S_{\phi K_S}$.

%%%%%%%%%%%%%%%%%%%%%%%%%%%%%%%%%%%%%%%%%%%%%%%%%%%%%%%%%%%%%%%%%%%%%%%%
\section{DNA-Flavour Test of $SSU(5)_{\rm RN}$}\label{sec:dna}
%%%%%%%%%%%%%%%%%%%%%%%%%%%%%%%%%%%%%%%%%%%%%%%%%%%%%%%%%%%%%%%%%%%%%%%%

The pattern of flavour violation predicted by specific NP model represents one of the
most powerful tools in the attempt to probe or to falsify the model in question. Motivated
by this consideration, in Ref.~\cite{Altmannshofer:2009ne} a ``DNA-Flavour Test'' has
been introduced with the aim of summarizing the potential size of deviations from the SM
results for the most interesting observables in a selection of SUSY and non-SUSY models.

In tab.~\ref{tab:DNA}, we extend such a ``DNA-Flavour Test'' to the $SSU(5)_{RN}$
model. We remind that we distinguish among large, moderate (but still visible) and
vanishingly small effects by three {\it red} stars, two {\it blue} stars and one
{\it black} star, respectively.

While we refer to Ref.~\cite{Altmannshofer:2009ne} for a detailed description of the
pattern of NP effects in various SUSY models, we want to comment here about one of
the most remarkable difference we found between the $SSU(5)_{RN}$ model and the
SUSY flavour models discussed in Ref.~\cite{Altmannshofer:2009ne}.

In fact, none of the models discussed in Ref.~\cite{Altmannshofer:2009ne} was able to
simultaneously account for the current data for $S_{\psi\phi}$ and $S_{\phi K_S}$, in
contrast to the $SSU(5)_{RN}$ model discussed here.

The reason for this can be traced back recalling that $S_{\phi K_S}$ receives the dominant
effects from gluino/squark penguins while $S_{\psi\phi}$ either from gluino/squark boxes
(at moderate/low $\tan\beta$ values) or from double Higgs penguins (at large $\tan\beta$).
However, only the moderate/low $\tan\beta$ solution can simultaneously account for an
enhancement of $S_{\psi\phi}$ and a suppression of $S_{\phi K_S}$ (relative to $S_{\psi K_S}$)
as required by the data.
Yet such effects are strongly constrained either by $D^0-\bar D^0$ mixing (in case of
Abelian flavour models) or by $K^0-\bar K^0$ mixing (in case of non-Abelian flavour models). Consequently, in this region of parameter space $S_{\psi\phi}$ cannot be large in these
models. In fact $S_{\psi\phi}$ receives in these models large values only at large
$\tan\beta$ where the sign of the correlation between $S_{\psi\phi}$ and $S_{\phi K_S}$
is found to be opposite to data ~\cite{Altmannshofer:2009ne}, that is $S_{\phi K_S}$ is
enhanced rather than suppressed when $S_{\psi\phi}$ is enhanced.

In contrast, the $SSU(5)_{RN}$ model predicts unobservable effects for $D^0-\bar D^0$
mixing while the NP effects in $K^0-\bar K^0$ are generally unrelated to those entering
$B^0_s-\bar B^0_s$ mixing and therefore the tight bounds from $\epsilon_K$ can be always
avoided. Therefore at moderate/low $\tan\beta$ the suppression of $S_{\phi K_S}$ and
simultaneous sizable enhancement of $S_{\psi\phi}$ can be obtained.

In this context let us recall that within the SM4, the SM with fourth sequential generation,
the correlation between $S_{\psi\phi}$ and in $S_{\phi K_S}$ is qualitatively similar to the
one found in the $SSU(5)_{RN}$ model, that is with increasing $S_{\psi\phi}$ the asymmetry
$S_{\phi K_S}$ decreases in accordance with the data~\cite{Hou:2005yb,Soni:2008bc,Buras:2010pi}.
However, in this model the absence of right-handed currents implies, in contrast to
$SSU(5)_{RN}$, that also $S_{\eta' K_S}$ decreases with increasing $S_{\psi\phi}$.

Finally, it has to be stressed that the ``DNA-Flavour Test'' table doesn't account for
possible correlations among observables. Therefore, since simultaneous large effects are
not always possible for certain sets of observables, it will be interesting to monitor
the changes in this table with improved experimental results.

%%%%%%%%%%%%%%%%%%%%%%%%%%%%%%%%%%%%%%%%%%%%%%%%%%%%%%%%%%%%%%%%%%%%%%%%
\definecolor{green1}{rgb}{0.06,0.66,0.06}
\definecolor{orange1}{rgb}{0.98,0.60,0.07}
\newcommand{\green}{{\color{green1}$\bigstar$}}
\newcommand{\orange}{{\color{orange1}\LARGE \protect\raisebox{-0.1em}{$\bullet$}}}
\newcommand{\red}{{\color{red}\small \protect\raisebox{-0.05em}{$\blacksquare$}}}
\newcommand{\three}{{\color{red}$\bigstar\bigstar\bigstar$}}
\newcommand{\two}{{\color{blue}$\bigstar\bigstar$}}
\newcommand{\one}{{\color{black}$\bigstar$}}
%%%%%%%%%%%%%%%%%%%%%%%%%%%%%%%%%%%%%%%%%%%%%%%%%%%%%%%%%%%%%%%%%%%%%%%%
%
\begin{table}[t]
\addtolength{\arraycolsep}{4pt}
\renewcommand{\arraystretch}{1.5}
\centering
\begin{tabular}{|l|c|c|c|c|c|c|c|}
\hline
 Observable & $SSU(5)_{\rm RN}$ model 
\\
\hline\hline
 $D^0-\bar D^0$ & \one
\\
\hline
$\epsilon_K$& \three
\\
\hline
$ S_{\psi K_S}$ & \one
\\
\hline
$ \gamma $  & \three
\\
\hline
$ R_t $  & \three
\\
\hline
$ S_{\psi\phi}$ & \three
\\
\hline\hline
$S_{\phi K_S}$ & \three \\
\hline
$S_{\eta^{\prime} K_S}$ & \two \\
\hline
$A_{\rm CP}\left(B\rightarrow X_s\gamma\right)$ & \one
\\
\hline
$A_{7,8}(B\to K^*\mu^+\mu^-)$ & \one
\\
\hline
$A_{9}(B\to K^*\mu^+\mu^-)$ & \one
\\
\hline
$B\to K^{(*)}\nu\bar\nu$  & \one
\\
\hline
$B_s\rightarrow\mu^+\mu^-$ & \three
\\
\hline
$K^+\rightarrow\pi^+\nu\bar\nu$ & \one
\\
\hline
$K_L\rightarrow\pi^0\nu\bar\nu$ & \one
\\
\hline
$\mu\rightarrow e\gamma$& \three \\
\hline
$\tau\rightarrow \mu\gamma$ & \three \\
\hline
$\mu + N\rightarrow e + N$& \three \\
\hline\hline
$d_n$& \three
\\
\hline
$d_{Hg}$& \three
\\
\hline
$d_e$& \three
\\
\hline
$\left(g-2\right)_\mu$& \three
\\
\hline
\end{tabular}
\renewcommand{\arraystretch}{1}
\caption{\small
``DNA'' of flavour physics effects for the $SSU(5)_{RN}$ model. \three\ signals large
effects, \two\ visible but small effects and \one\ implies vanishingly small effects.}
\label{tab:DNA}
\end{table}

\section{Conclusions}

Despite of the remarkable agreement of flavour data with the SM predictions in the $K$
and $B_d$ systems, a closer look at the data might indicate some tensions especially in
CP violating observables. In particular, the most recent UT analyses show some tensions
at the level of $(2-3)\sigma$~\cite{SL,Buras:2008nn,Lenz:2010gu,Bevan:2010gi} and recent
messages from the Tevatron seem to hint the presence of new sources of CPV entering the
$B^{0}_s$ systems~\cite{Aaltonen:2007he,Abazov:2010hv,Abazov:2008fj}.

Motivated by the above facts, in the present work, we have analyzed the low energy implications of a supersymmetric $SU(5)$ GUT scenario with right-handed neutrinos~\cite{Hisano:1997tc} ($SSU(5)_{RN}$) accounting for the neutrino
masses and mixing angles by means of a type-I see-saw mechanism~\cite{seesaw}.

Since supersymmetric Grand Unified theories generally predict FCNC and CP violating
processes to occur both in the leptonic and hadronic sectors, we have performed
an extensive study of FCNC and CP Violation in both sectors, analyzing possible
hadron/lepton correlations among observables. In particular, we have monitored
the low energy consequences implied by the solutions to the above tensions.

However, within the $SSU(5)_{RN}$ model, it is not possible to link model
independently different flavour transitions like $s\to d$ and $b\to s$. In fact,
the neutrino Yukawa couplings, which regulate the size of the flavour violation
both in the hadronic and leptonic sectors, are unknown. Therefore, we have analyzed 
the phenomenology related to $s\to d$ and $b\to s$ transitions separately.

The main results of our study of the  $s\to d$ transitions and their correlation
with $\mu\to e$ transitions are

\begin{itemize}
\item Sizable SUSY effects in $\epsilon_{K}$, that might be desirable to solve
the UT anomaly, generally imply a lower bound for ${\rm BR}(\mu\to e\gamma)$ in
the reach of the MEG experiment. Furthermore, the simultaneous requirement of an
explanation for both the $(g-2)_\mu$ and the UT anomalies would typically imply
${\rm BR}(\mu\to e\gamma)\geq 10^{-12}$.
\item The requirement of sizable non-standard effects in $\epsilon^{SUSY}_{K}$ always
implies large values for the electron and neutron EDMs, in the reach of the planned
experimental resolutions.
\end{itemize}

The main results of our study of the $b\to s$ transitions and of their correlations with
$\tau\to\mu$ transitions are

\begin{itemize}
\item Non-standard values for $S_{\psi\phi}$ imply a lower bound for ${\rm BR}(\tau\to\mu\gamma)$
within the SuperB reach. However, the $(g-2)_\mu$ anomaly can be solved only for large $\tan\beta$
values where we find $|S_{\psi\phi}|\leq 0.2$ for $\Delta a^{\rm SUSY}_{\mu}\gtrsim 1\times 10^{-9}$ while being still compatible with the constraints from ${\rm BR}(\tau\to\mu\gamma)$.
\item The UT anomaly can be solved by means of negative NP effects in $\Delta M_d/\Delta M_s$
which, in turn, also indirectly enhance $\epsilon_K$ via the increased value of $R_t$.
This scenario implies a lower bound for ${\rm BR}(\tau\to\mu\gamma)$ within the SuperB
reach and large values for the angle $\gamma$ and it will be probed or falsified quite
soon at the LHCb.
\item Both ${\rm BR}(B_s\to\mu^+\mu^-)$ and ${\rm BR}(B_d\to\mu^+\mu^-)$ can reach
large non-standard values. However, sizable departures from the MFV prediction
${\rm BR}(B_s\to\mu^+\mu^-)/{\rm BR}(B_d\to\mu^+\mu^-)\approx|V_{ts}/V_{td}|^2$ would
imply large values for ${\rm BR}(\tau\to\mu\gamma)$, well within the SuperB reach.
\item The dileptonic asymmetry $A^{b}_{\text{SL}}$ can sizably depart from the SM
expectations but the large value reported by the Tevatron~\cite{Abazov:2010hv} cannot
be accounted for within the $SSU(5)_{RN}$ model. In particular, we find that
$A^{b}_{\text{SL}} \approx 0.5~A^{s}_{\text{SL}}$ since $A^{d}_{\text{SL}}$ remains
SM-like.
\item The asymmetry $S_{\phi K_S}$ can sizably depart from the SM expectations
and it turns out to be correlated with $S_{\psi\phi}$. In particular, it is possible
to simultaneously account for an enhancement of $S_{\psi\phi}$ and a suppression of
$S_{\phi K_S}$ (relative to $S_{\psi K_S}$) as required by the data. This is in
contrast to the SUSY flavour models discussed in Ref.~\cite{Altmannshofer:2009ne}.
Moreover, the asymmetries $S_{\eta^{\prime} K_S}$ and $S_{\phi K_S}$ exhibit opposite
deviations with respect to the SM expectations.

\end{itemize}

Finally, we provided a ``DNA-Flavour Test'' (proposed in Ref.~\cite{Altmannshofer:2009ne})
for the $SSU(5)_{RN}$ model, with the aim of showing a tool to distinguish between
NP scenarios, once additional data on flavour changing processes become available.

As shown in tab.~\ref{tab:DNA}, further important predictions of the $SSU(5)_{RN}$ model
are that i) $S_{\psi K_S}$ remains SM-like to a very good extent (therefore, the solution of
the UT anomaly by means of CPV effects in $b\to d$ mixing is not possible), ii) CPV effects
in $D^{0}-\bar{D}^{0}$ are negligibly small, and iii) $BR(K^{0}_{L}\to\pi^{0}\nu\bar{\nu})$
and $BR(K^{+}\to \pi^{+}\nu\bar{\nu})$ also remain SM-like.

In conclusion, the above results show the richness which is present in flavour physics
once we embed a GUT group within a gravity mediated SUSY breaking scenario. It will be
exciting to monitor upcoming results from the Tevatron, LHC(b), the MEG experiment at
PSI and SuperB machines to establish whether some patterns of deviations from the SM 
expectations we have pointed out in this work are at work or not.

The interplay of all these efforts with the direct searches for NP will be most exciting.

\vspace*{20pt}

\noindent
\textit{Acknowledgments:}
AJB would like to thank the Particle Theory Institute of Vienna University for its
hospitality during the final steps of this work.
This work has been supported in part by the Cluster of Excellence ``Origin and Structure
of the Universe'' and by the German Bundesministerium f{\"u}r Bildung und Forschung under
contract 05H09WOE.

\end{document}